\documentclass[a4paper,onecolumn,showpacs,notitlepage,floatfix,nofootinbib,superscriptaddress,prd]{revtex4-1}
\usepackage{graphicx}
\usepackage{amsfonts}
\usepackage{amsmath}
\usepackage{hyperref}
\usepackage{float}
\usepackage{amssymb,amsbsy}
\usepackage{epstopdf}
\usepackage{color}
\usepackage{oldgerm}

\begin{document}

\title{Non-resistive dissipative magnetohydrodynamics from the Boltzmann equation in the 14-moment approximation}
\author{Gabriel S.\ Denicol}
\affiliation{Instituto de F\'isica, Universidade Federal Fluminense, UFF, Niter\'oi, 24210-346, RJ, Brazil}
\author{Xu-Guang Huang}
\affiliation{Physics Department and Center for Particle Physics and Field Theory, 
Fudan University, Shanghai 200433, China}
\affiliation{Key Laboratory of Nuclear Physics and Ion-beam Application
(MOE), Fudan University, Shanghai 200433, China}
\author{Etele Moln\'ar}
\affiliation{Institut f\"ur Theoretische Physik, 
Johann Wolfgang Goethe--Universit\"at,
Max-von-Laue-Str.\ 1, D--60438 Frankfurt am Main, Germany} 
\affiliation{Institute of Physics and Technology, University of Bergen, Allegaten 55, 5007 Bergen, Norway}
\author{Gustavo M.\ Monteiro}
\affiliation{Instituto de F\'isica Gleb Wataghin, Universidade Estadual de Campinas-UNICAMP, 13083-859 Campinas, 
SP, Brazil}
\author{Harri Niemi}
\affiliation{Institut f\"ur Theoretische Physik, 
Johann Wolfgang Goethe--Universit\"at,
Max-von-Laue-Str.\ 1, D--60438 Frankfurt am Main, Germany} 
\affiliation{Department of Physics, University of Jyv\"askyl\"a, P.O.\ Box 35, FI-40014 University of Jyv\"askyl\"a, Finland}
\affiliation{Helsinki Institute of Physics, P.O.\ Box 64, FI-00014 University of Helsinki, Finland}
\author{Jorge Noronha}
\affiliation{Instituto de F\'isica, Universidade de S\~ao Paulo, Rua do Mat\~ao, 1371,
Butant\~a, 05508-090, S\~ao Paulo, SP, Brazil}
\author{Dirk H.\ Rischke}
\affiliation{Institut f\"ur Theoretische Physik, 
Johann Wolfgang Goethe--Universit\"at,
Max-von-Laue-Str.\ 1, D--60438 Frankfurt am Main, Germany} 
\affiliation{Department of
Modern Physics, University of Science and Technology of China, Hefei, Anhui 230026, China}
\author{Qun Wang}
\affiliation{Department of
Modern Physics, University of Science and Technology of China, Hefei, Anhui 230026, China}

\begin{abstract}
We derive the equations of motion of relativistic, non-resistive, second-order dissipative magnetohydrodynamics
from the Boltzmann equation using the method of moments. We assume the fluid to be composed of
a single type of point-like particles with vanishing dipole moment or spin, so that the fluid has vanishing 
magnetization and polarization.
In a first approximation, we assume the fluid to be non-resistive, which allows to express the electric field in terms of
the magnetic field. We derive equations of motion for the irreducible moments of the deviation of the
single-particle distribution function from local thermodynamical equilibrium. We analyze the Navier-Stokes limit
of these equations, reproducing previous results for the structure of the first-order transport coefficients. 
Finally, we truncate the system of equations
for the irreducible moments using the 14-moment approximation, deriving the equations of motion of
relativistic, non-resistive, second-order dissipative magnetohydrodynamics. We also  
give expressions for the new transport coefficients appearing due to the coupling of the magnetic field to the 
dissipative quantities.

\end{abstract}

\date{\today }
\maketitle

%%%

\section{Introduction}

The success of relativistic fluid dynamics in describing the evolution
of high-energy heavy-ion collisions~\cite{Heinz:2013th} and the existence
of very large magnetic fields in these collisions~\cite{Skokov:2009qp,Deng:2012pc,Tuchin:2013apa,Bzdak:2011yy} has
generated a lot of interest in observing the effects of the magnetic
field on the fluid-dynamical evolution in these systems. The generic
framework that couples the electromagnetic field to the dynamics of a fluid
is referred to as magnetohydrodynamics \cite{degroot,Rezzolla_book:2013}.
There are several works where the effect of electromagnetic fields on the dynamics
of heavy-ion collisions have been studied [for a review, see Ref.\ \cite{Huang:2015oca} and refs.\ therein], 
but so far they have been mostly based on the 
non-resistive, non-dissipative formulation of relativistic magnetohydrodynamics. However,
dissipation plays an important role in understanding the dynamics of
heavy-ion collisions and in particular in explaining the magnitude of the observed collective flow
[for a review, see Ref.\ \cite{Heinz:2013th} and refs.\ therein]. 
Thus, it is essential to develop a relativistic formulation of dissipative magnetohydrodynamics. 

In principle, the most simple dissipative fluid-dynamical theory is a
relativistic generalization of Navier-Stokes theory, where the
dissipative quantities, bulk viscous pressure, diffusion currents, and
shear-stress tensor, are proportional to the gradients of the flow field
and of thermodynamical quantities. In the absence of a magnetic field, the
constants of proportionality are three scalar transport coefficients: bulk viscosity,
diffusion constant, and shear viscosity. A magnetic field breaks the isotropy of space,
introducing several new transport 
coefficients~\cite{Huang:2009ue,Huang:2011dc,Critelli:2014kra,Finazzo:2016mhm,Hernandez:2017mch}, 
which assume different values
in the direction of the magnetic field and in the direction orthogonal to it. 
The relativistic generalization of Navier-Stokes theory is, however, known to be acausal \cite{pichon} and,
at least, linearly unstable~\cite{Hiscock,Denicol:2008ha,Pu:2009fj},
rendering it ill-suited for practical
use. Without the magnetic field, these problems were cured by the
causal and stable ``second-order'' formalism of Israel and Stewart \cite{Israel:1976tn,Stewart:1977,Israel:1979wp}. 
Israel-Stewart theory can be derived by starting from the relativistic
Boltzmann equation employing the so-called 14-moment
approximation~\cite{Stewart:1977,Israel:1979wp}, and the success of fluid dynamics
in describing the dynamics of heavy-ion collisions is based on this formalism. 

In this paper, we follow the same line of reasoning as Israel and Stewart, and
derive a relativistic causal theory of second-order dissipative
magnetohydrodynamics from the relativistic Boltzmann equation coupled
to an electromagnetic field. As in the original formulation by Israel
and Stewart, we restrict ourselves to a single-component system of
spinless particles undergoing binary elastic collisions 
and use the 14-moment approximation in the framework developed in 
Refs.\ \cite{Denicol:2010xn,Denicol:2012cn,Denicol:2012es}. In a first step,
we assume the fluid to have infinite electric conductivity (or zero resistivity), which allows 
to replace the electric field by the magnetic field and considerably simplifies the equations of motion. 
We remark that the assumption of infinite electric conductivity is an idealization which is hard 
(if not impossible) to realize in 
systems whose microscopic dynamics is described by the Boltzmann equation:
the electric conductivity is a transport coefficient which is
proportional to the mean free path of the particles, such as all other
transport coefficients appearing in dissipative fluid dynamics, and thus should be of the same order of
magnitude as the latter. Nevertheless, as non-resistive magnetohydrodynamics is a theory which finds
widespread applications, we decided to first discuss the simpler case of a non-resistive (albeit dissipative) fluid.
The generalization towards systems with finite conductivity will be addressed in a subsequent paper.

Let us add a few remarks on the length scales entering our discussion:
(i) The Boltzmann equation is derived under the assumption that the collision term in this equation is local,
implying that the mean free path $\lambda_{\rm mfp}$ between
collisions is much larger than the typical interaction length $\sqrt{\sigma/\pi}$,
where $\sigma$ is the binary-collision cross section. 
(ii) The magnetic field leads to cyclotron motion of the charged particles. The curvature of the particle
trajectories is given by the inverse Larmor radius $R_L^{-1} = \textswab{q}B/k_\perp$, where
$\textswab{q}$ is the electric charge of the particles and
$k_\perp$ is the momentum of the particle transverse to the direction of the magnetic induction field
$\mathbf{B}$, which has magnitude  $B= |\mathbf{B}|$ (in the following the magnetic induction field is
in a simplifying, but somewhat incorrect, manner referred to as ``magnetic field''). 
In our discussion we will assume that the magnetic field is sufficiently weak so that
we can neglect the Landau quantization of the cyclotron motion. This implies
that the thermal energy $\sim T$, where $T$ is the temperature,
is much larger than the cyclotron frequency $ \sim \sqrt{\textswab{q}B}$.
In other words, the thermal wavelength $\beta_0 \equiv 1/T \ll R_T$, where $R_T \equiv (\textswab{q}B \beta_0)^{-1}$
is the Larmor radius of a particle with transverse momentum $k_\perp = T$.
In the following, we refer to $R_T$ as ``thermal Larmor radius''.
Note that this condition does not necessarily imply that the magnetic field is weak in absolute magnitude, it only requires
that the temperature of the system is sufficiently large, such that $T^2 \gg\textswab{q}B$.
While our discussion is valid when $\lambda_{\rm mfp} \gg \sqrt{\sigma/\pi}$ and
$R_T \gg \beta_0$, there is \emph{a priori} no constraint on the ratio $\xi_B \equiv \lambda_{\rm mfp}/R_T
= \textswab{q}B \beta_0\lambda_{\rm mfp}$ \cite{Li:2017tgi}, as long as the first two inequalities are fulfilled.

This paper is organized as follows. In Sec.~\ref{sec:eom} we
review the structure of the equations of motion of magnetohydrodynamics, i.e., the evolution equations
for energy and momentum coupled to Maxwell's equations for the electromagnetic fields. 
In Sec.~\ref{sec:non_diss_eom} we
present the magnetohydrodynamic equations of motion for the non-resistive,
non-dissipative fluid. In Sec.~\ref{sec:diss_eom_moments} we recall
the method of moments and derive the equations
of motion for the moments of the deviation of the single-particle distribution function from local
thermodynamical equilibrium in the presence of a magnetic field. In Sec.~\ref{sec:diss_eom_NS} we
show how the Navier-Stokes limit arises from the moment expansion. Finally, 
in Sec.~\ref{sec:diss_eom_2ndorder} we derive the main
result of this paper, the equations of
motion for non-resistive, second-order dissipative magnetohydrodynamics.
Section \ref{conclusions} concludes this work with a summary of the results and an outlook
to future work.

We adopt natural Heaviside-Lorentz units, $\hbar =c=\epsilon_0 = \mu_0 = k_{B}=1$. Our convention for the 
metric tensor is $g^{\mu \nu }=\text{diag}(1,-1,-1,-1)$. The fluid four-velocity is 
$u^{\mu}\left( t,\mathbf{x}\right) =\gamma \left( 1,\mathbf{v}\right)^T $, with 
$\gamma =(1-\mathbf{v}^2)^{-1/2}$, leading to the normalization
$u^{\mu }u_{\mu }\equiv 1$. In the local rest (LR) frame of the
fluid, $u_{LR}^{\mu }=\left(1,\mathbf{0}\right)^T $. 
The four-momentum $k^\mu$ of particles is normalized to their rest mass $m_{0}$,
$k^{\mu }k_{\mu }=m_{0}^{2}$. The rank-two projection operator onto the
three-space orthogonal to $u^{\mu }$ is $\Delta ^{\mu \nu }=g^{\mu \nu }-u^{\mu }u^{\nu }$. 
For a four-vector $A^\mu$, we define its projection onto the three-dimensional subspace orthogonal to $u^\mu$ as
$A^{\left\langle \mu \right\rangle} \equiv \Delta^\mu_\nu A^\nu$.
The rank-four projection operator is defined as $\Delta _{\alpha \beta }^{\mu \nu }
=\frac{1}{2}\left( \Delta _{\alpha}^{\mu }\Delta _{\beta }^{\nu }+\Delta _{\beta }^{\mu }\Delta _{\alpha}^{\nu }\right) 
-\frac{1}{3}\Delta ^{\mu \nu }\Delta _{\alpha \beta }$, which is symmetric and traceless.
For a rank-two tensor, we define the symmetric, traceless projection onto the
three-space orthogonal to $u^\mu$ as
$A^{\left\langle \mu \nu \right\rangle} \equiv \Delta _{\alpha \beta }^{\mu \nu } A^{\alpha \beta}$.
Our convention and useful relations for the rank-four Levi-Civit\`a tensor 
$\epsilon^{\mu \nu\alpha \beta}$ are given in the Appendix.

\section{Equations of motion of magnetohydrodynamics}
\label{sec:eom}

\subsection{Maxwell's equations and energy-momentum tensor of the electromagnetic field} 

In a relativistically covariant formulation of electrodynamics, the electric field vector
$\mathbf{E}$ and the magnetic field vector $\mathbf{B}$
constitute the components of the Faraday tensor $F^{\mu \nu }$. The latter is an antisymmetric (and hence traceless) 
rank-two tensor (and thus has six independent components, corresponding to the six components
of $\mathbf{E}$ and $\mathbf{B}$). Without loss of generality it can be decomposed
with respect to the fluid velocity as \cite{Cercignani_book, Barrow:2006ch}
\begin{equation}
F^{\mu \nu } \equiv E^{\mu }u^{\nu }-E^{\nu }u^{\mu }+\epsilon ^{\mu \nu \alpha \beta}u_{\alpha }B_{\beta }\;,  \label{F_munu}
\end{equation}
while its Hodge dual is 
\begin{equation}
\tilde{F}^{\mu \nu }\equiv \frac{1}{2}\epsilon ^{\mu \nu \alpha \beta}F_{\alpha \beta }
=B^{\mu }u^{\nu }-B^{\nu }u^{\mu }-\epsilon ^{\mu \nu
\alpha \beta }u_{\alpha }E_{\beta }\;.  \label{F_munu_dual}
\end{equation}
Here we defined the electric field four-vector $E^{\mu }\equiv F^{\mu \nu }u_{\nu}$ and the
magnetic field four-vector $B^{\mu }\equiv \tilde{F}^{\mu \nu }u_{\nu }
=\frac{1}{2}\epsilon^{\mu \nu \alpha \beta }F_{\alpha \beta }u_{\nu }$. Using the antisymmetry of the Faraday tensor
and the rank-four Levi-Civit\`a tensor, one readily realizes that $E^\mu$ and $B^\mu$ are orthogonal
to the fluid velocity, $E^{\mu }u_{\mu }=0$ and $B^{\mu }u_{\mu }=0$. Moreover, in the local rest frame
of the fluid, they coincide with
the usual electric and magnetic fields, i.e., $E^{\mu }_{LR}=\left( 0,\mathbf{E}\right)^T $ and $B^{\mu}_{LR}=
\left( 0,\mathbf{B}\right)^T $, with $\mathrm{E}^{i}=F^{i0}$ and 
$\mathrm{B}^{i}=-\frac{1}{2}\epsilon ^{ijk}F_{jk}$. The electric field
is a polar vector, while the magnetic field is an axial vector dual to $F_{jk}$.

The evolution of the electric and magnetic fields are given by Maxwell's equations,
\begin{eqnarray}
\partial _{\mu }F^{\mu \nu } &=&\textswab{J}^{\nu }\;,  \label{Maxwell_inhom} \\
\partial _{\mu }\tilde{F}^{\mu \nu } &=&0\; ,
\label{Maxwell_hom}
\end{eqnarray}
where the electric charge four-current $\textswab{J}^{\nu }$ serves as source for the electromagnetic field. It 
can be tensor-decomposed with respect to the fluid velocity \cite{Barrow:2006ch, Bekenstein_78},
\begin{equation}
\textswab{J}^{\mu} = \textswab{n} u^{\mu } + \textswab{V}^{\mu }\;,  \label{J_mu_em}
\end{equation}
where $\textswab{n} = u_{\mu}\textswab{J}^\mu $ is the charge density in the local rest frame of the fluid
and $\textswab{V}^\mu \equiv \Delta^{\mu}_\nu \textswab{J}^{\nu} $ is the charge diffusion four-current. The solution of Eqs.\ (\ref{Maxwell_inhom}) and (\ref{Maxwell_hom}) determines
the electromagnetic fields as functionals of $\textswab{J}^\mu$.

For non-polarizable, non-magnetizable fluids the electromagnetic stress-energy tensor is given 
by \cite{Cercignani_book,Israel:1978up} 
\begin{equation}
T_{em}^{\mu \nu }=-F^{\mu \lambda }F_{\left.{}\right. \lambda }^{\nu }
+\frac{1}{4}g^{\mu \nu }F^{\alpha \beta }F_{\alpha
\beta }\; .  \label{T_munu_em}
\end{equation}
Using Maxwell's equations (\ref{Maxwell_inhom}), (\ref{Maxwell_hom}) one can show that
\begin{equation}
\partial _{\mu }T_{em}^{\mu \nu } =-F^{\nu \lambda }\textswab{J}_{\lambda }\;.
\label{d_mu_T_munu_em}
\end{equation}

\subsection{Particle four-current and energy-momentum tensor of the fluid}
\label{sec:IIB}

For particles without a microscopic dipole moment or spin the canonical momentum coincides with
the kinetic momentum \cite{Israel:1978up}. Then, the particle four-current and energy-momentum
tensor of the fluid are simply given by
\begin{eqnarray}
N_{f}^{\mu } &\equiv &\left\langle k^{\mu }\right\rangle\;,
\label{kinetic:N_mu} \\
T_{f}^{\mu \nu } &\equiv &\left\langle k^{\mu }k^{\nu }\right\rangle \;.  \label{kinetic:T_munu}
\end{eqnarray}
Here, 
\begin{equation} \label{average}
\left\langle \cdots \right\rangle \equiv \int dK\, \cdots\, f_{\mathbf{k}}\;,
\end{equation}
with $f_{\mathbf{k}}$ being the single-particle distribution function and
$dK \equiv g\, d^3\mathbf{k}/[(2 \pi)^3 k^0]$ being the Lorentz-invariant measure in momentum space,
where $g$ is the degeneracy factor due to internal degrees of freedom (note, however, 
that the spin degeneracy is $2J+1 = 1$, since we consider spin-zero particles), and $k^0 = \sqrt{\mathbf{k}^2 + m_0^2}$
is the on-shell energy.

The particle four-current and the energy-momentum tensor can be tensor-decomposed with respect to the fluid velocity,
\begin{eqnarray}
N_{f}^{\mu } & =&n_fu^{\mu }+V_f^{\mu }\;,\\
T_{f}^{\mu \nu } &=& \varepsilon u^{\mu }u^{\nu }-P\Delta ^{\mu \nu }+W^{\mu }u^{ \nu } + W^\nu u^\mu +\pi ^{\mu \nu }\;, 
\end{eqnarray}
where the particle density $n_f$, the
energy density $\varepsilon$, and the isotropic pressure $P$ are defined as 
\begin{eqnarray}
n_f &\equiv &N_{f}^{\mu }u_{\mu }=\left\langle E_{\mathbf{k}}\right\rangle\; ,
\label{kinetic:n} \\
\varepsilon &\equiv &T_{f}^{\mu \nu }u_{\mu }u_{\nu }=\left\langle E_{\mathbf{k}}^{2}\right\rangle \;,  \label{kinetic:e} \\
P &\equiv &-\frac{1}{3}T_{f}^{\mu \nu }\Delta _{\mu \nu }
=-\frac{1}{3}\left\langle \Delta ^{\mu \nu }k_{\mu }k_{\nu }\right\rangle\; ,
\label{kinetic:P}
\end{eqnarray}
with $E_{\mathbf{k}}=k^{\mu }u_{\mu }$ being the energy of a particle in the local rest frame of the fluid.
The particle and energy-momentum diffusion currents orthogonal to the flow
velocity are 
\begin{eqnarray}
V_f^{\mu } &\equiv &\Delta _{\nu }^{\mu }N_{f}^{\nu }=\left\langle
k^{\left\langle \mu \right\rangle }\right\rangle\; ,  \label{kinetic:V_mu} \\
W^{\mu } &\equiv &\Delta _{\alpha }^{\mu }T_{f}^{\alpha \beta }u_{\beta}
=\left\langle E_{\mathbf{k}}k^{\left\langle \mu \right\rangle}\right\rangle\;,  \label{kinetic:W_mu}
\end{eqnarray}
respectively, while the shear-stress tensor is 
\begin{equation}
\pi^{\mu \nu }\equiv \Delta _{\alpha \beta }^{\mu \nu }T_{f}^{\alpha \beta}
=\left\langle k^{\left\langle \mu \right. }k^{\left. \nu \right\rangle}\right\rangle\; . \label{kinetic:pi_munu}
\end{equation}

For a single-component fluid, the electric charge and particle four-currents are related by
\begin{equation}
\textswab{J}_f^{\mu } \equiv \textswab{q} N_f^\mu 
= \textswab{n}_f u^{\mu }+\textswab{V}_f^{\mu }\;,  \label{J_f_mu}
\end{equation}
where $\textswab{n}_f \equiv u_\nu \textswab{J}_f^\nu \equiv \textswab{q} u_\nu N_f^\nu = \textswab{q} n_f$ 
is the charge density in the local rest frame and
$\textswab{V}_f^\mu \equiv \Delta^{\mu}_\nu \textswab{J}_f^\nu \equiv \textswab{q} \Delta^\mu_\nu N_f^\nu 
= \textswab{q} V_f^\mu$ is the charge diffusion current. To leading order, the charge diffusion current
is equal to the Ohmic induction current,
$\textswab{q} V_f^\mu \simeq \textswab{J}_{ind}^{\mu}=\sigma _{E}E^{\mu }$.

%Hence, the five equations of motion of magnetohydrodynamics,
%Eqs.\ (\ref{T_munucons}), (\ref{J_f_mucons}) contain 14 unknowns, the components of 
The components of $\textswab{J}_f^\mu$ 
and $T^{\mu \nu}_f$ contain 14 unknowns, 
or equivalently, for a given four-vector field $u^\mu$ the three scalar quantities $\textswab{n}_f, \,\varepsilon, \,P$, 
the two times three (equals six) independent components of
$\textswab{V}_f^\mu$ and $W^\mu,$ and the five independent components of $\pi^{\mu \nu}$.
If the fluid velocity is a dynamical quantity, this would add another three
unknowns (the three independent components of $u^\mu$). However, the fluid velocity can be chosen
to be proportional to the charge four-current, which eliminates the charge diffusion current $\textswab{V}_f^\mu$ 
[the so-called Eckart frame \cite{Eckart:1940te}], or to be proportional to the flow of energy, which eliminates the 
energy-momentum
diffusion current $W^\mu$ [the so-called Landau frame \cite{Landau_book}].
%This still leaves 14 unknowns, 
%but only five equations of motion. Thus, in order to close the
%system of equations of motion, we need to make approximations, which we will discuss in the following section.

Let us assume that the only charge current in the system is that of the fluid,
$\textswab{J}^\mu \equiv \textswab{J}^\mu_f$. If we project Maxwell's equation (\ref{Maxwell_inhom}) onto 
$u_{\nu}$ and use Eqs.\ (\ref{F_munu}) and (\ref{J_mu_em}), we obtain, 
\begin{equation}
\nabla _{\mu }E^{\mu }+2\omega _{\mu }B^{\mu }=\textswab{n}_f\;,  \label{Gauss}
\end{equation}
where we introduced the three-space gradient $\nabla _{\mu }\equiv 
\Delta_{\mu }^{\alpha }\partial _{\alpha }$ and the vorticity four-vector 
\begin{equation}
\omega ^{\mu }=\frac{1}{2}\epsilon ^{\mu \nu \alpha \beta }u_{\nu }\partial_{\alpha }u_{\beta }\;.
\end{equation}
In the following, we want to consider the non-resistive limit, i.e., the electric
conductivity $\sigma _{E}\rightarrow \infty $. In this limit, the Ohmic conduction current
$\textswab{J}_{ind}^\mu$ would diverge, unless we demand that 
$E^{\mu }=0$, or $\mathbf{E} = - \mathbf{v} \times \mathbf{B}$, so that $\textswab{J}_{ind}^\mu
\simeq \textswab{V}_f^\mu$ remains finite.
However, if $E^\mu =0$, we observe that the
charge density of the fluid (and thus, for our single-component system, the
particle density of the fluid) assumes a value which is uniquely
determined by the scalar product of the magnetic field four-vector and the
fluid vorticity, $\textswab{n}_f=\textswab{q} n_f = 2\omega _{\mu }B^{\mu }$,\footnote{%
Amusingly, the corresponding term is of the same structure as the
spin-vorticity coupling term discussed in Ref.\ \cite{Hattori:2016njk}, but
the coefficient assumes a different value, since in that case it is
determined by spin-1/2 fermions in the lowest Landau level, while here we
deal with spinless particles and neglect Landau quantization.} and is no
longer an independent variable. Projecting Eq.\ (\ref{Maxwell_inhom}) with 
$\Delta _{\nu }^{\alpha }$, similar arguments apply to the charge diffusion
current $\textswab{V}^{\nu }_f = \textswab{q} V_f^\nu$. 

On the other hand, in dissipative fluid dynamics $n_{f}$ and $V_{f}^{\nu }$
are traditionally considered as four (out of 14) independent variables. In
order to maintain this feature, we introduce an external current, 
$\textswab{J}_{ext}^{\mu }$, such that the total charge current (\ref{J_mu_em}) reads 
\begin{equation}
\textswab{J}^{\mu }=\textswab{J}_{ext}^{\mu }+\textswab{J}_{f}^{\mu}\;.
\label{J_mu_ext}
\end{equation}
Then, $n_{f}$ and $V_{f}^{\nu }$ become independent variables to be determined by
the equations of motion for the fluid.
In this case, our derivation of
dissipative magnetohydrodynamics can be formulated in close analogy to the
one of ordinary dissipative fluid dynamics for single-component systems.
Note that the introduction of an external current does not affect our argument that $E^\mu$ 
must vanish in the limit of infinite conductivity.

\subsection{Equations of motion of magnetohydrodynamics}

The total energy-momentum tensor of the system is
\begin{equation}
T^{\mu \nu }=T_{em}^{\mu \nu }+T_{f}^{\mu\nu } \; .  \label{T_munu}
\end{equation}
Note that the separation of $T^{\mu \nu}$ into $T_{em}^{\mu \nu}$ and $T_f^{\mu \nu}$ is not
unique in the case of polarizable, magnetizable fluids \cite{Israel:1978up}. This problem is absent here, as
we consider a non-polarizable, non-magnetizable fluid. 

While the charge current of the fluid is conserved,
\begin{equation}\label{J_f_mucons} 
\partial_{\mu} \textswab{J}_f^{\mu } = 0 \;, 
\end{equation}
the total energy and momentum of the system are not, as the external charge current induces electromagnetic fields
and thus feeds energy and momentum into the system. In analogy to Eq.\ (\ref{d_mu_T_munu_em}) we have
\begin{equation} \label{T_munucons}
\partial _{\mu }T^{\mu \nu } = - F^{\nu \lambda } \textswab{J}_{ext,\lambda }\;.
\end{equation}
With Eq.\ (\ref{J_mu_ext}), Eq.\ (\ref{d_mu_T_munu_em}) reads
\begin{equation} \label{d_mu_T_munu_em2}
\partial _{\mu }T_{em}^{\mu \nu } = - F^{\nu \lambda } 
\left( \textswab{J}_{ext,\lambda } + \textswab{J}_{f,\lambda }\right)\;, 
\end{equation}
and with Eq.\ (\ref{T_munu}) we can derive from Eq.\ (\ref{T_munucons}) an equation of motion for the 
energy-momentum tensor of the fluid,
\begin{equation} \label{d_mu_T_munu_f}
\partial _{\mu }T_{f}^{\mu \nu } = F^{\nu \lambda }\textswab{J}_{f,\lambda }\;.
\end{equation}
Equations (\ref{J_f_mucons}), (\ref{d_mu_T_munu_em2}), and (\ref{d_mu_T_munu_f}) constitute the
equations of motion of magnetohydrodynamics. While the energy and momentum of the electromagnetic fields
change on account of the external charge current as well as the internal charge current of the particles in the fluid,
Eq.\ (\ref{d_mu_T_munu_em2}), the energy and momentum of the fluid change only on account of the 
Lorentz force exerted on the charged particles within the fluid by the electromagnetic fields, Eq.\ (\ref{d_mu_T_munu_f}).
In general, neither energy and momentum of the electromagnetic fields nor that of the fluid are conserved separately.
The total energy and momentum are only conserved in the absence of an external charge current, 
$\textswab{J}_{ext, \lambda} =0$, so that Eq.\ (\ref{T_munucons}) becomes $\partial _{\mu }T^{\mu \nu } =0$.

%While the electromagnetic fields influence the motion of the fluid via the Lorentz force exerted on
%$\textswab{J}_{f,\lambda }$, the latter influences the electromagnetic fields via the right-hand side of 
%Eq.\ (\ref{d_mu_T_munu_em2}), thus providing a backreaction of the fluid on the electromagnetic fields. 
%However, if the external charge current is considerably larger than the charge current of the fluid, 
%$|\textswab{J}^{\nu }_{ext}| \gg |\textswab{J}^{\nu }_{f}|$, the first
%Maxwell equation (\ref{Maxwell_inhom}) may be written as
%$\partial_{\mu} F^{\mu\nu} = \textswab{J}^{\nu }_{ext} + \textswab{J}^{\nu }_{f} \simeq \textswab{J}_{ext}^{\nu }$.
%In this case, the dynamics of the fields becomes (approximately) independent of the charge current of the fluid, and
%Eq.\ (\ref{d_mu_T_munu_em2}) reads
%\begin{equation} \label{limit_external}
%\partial _{\mu }T_{em}^{\mu \nu } \simeq - F^{\nu \lambda } 
%\textswab{J}_{ext,\lambda }\;. 
%\end{equation}
%On the other hand, the fields always influence the fluid
%motion through the Lorentz force exerted on the fluid charge current $\textswab{J}_{f,\lambda}$, which therefore 
%cannot be neglected in Eq.\ (\ref{d_mu_T_munu_f}). However, since in this case the feedback from the
%fluid motion to the evolution of the fields is neglected, the latter can be regarded as external (i.e.,
%being generated exclusively by the external current $\textswab{J}_{ext}^{\nu }$).

\section{Non-resistive, non-dissipative magnetohydrodynamics}
\label{sec:non_diss_eom}

\subsection{Assumption of non-resistivity}
\label{sec:IIIA}

The charge four-current induced by an electric field is $\textswab{J}_{ind}^\mu = \sigma_E E^\mu$. 
A widely used approximation in applications of magnetohydrodynamics is the assumption that the fluid is non-resistive,
i.e., ideally conducting, such that $\sigma_E \rightarrow \infty$. Then, as already stated, in order to have a
finite induced charge current $\textswab{J}_{ind}^\mu$ one has to demand $E^\mu \rightarrow 0$.
From this condition follows that, in an arbitrary frame, $\mathbf{E} = - \mathbf{v} \times \mathbf{B}$, such that 
the electric field can be eliminated from the equations of motion. 

An ideally conducting fluid implies an infinite mean free path of charged particles, i.e., the free-streaming limit.
However, in this paper we aim at deriving dissipative magnetohydrodynamics from an expansion around 
local thermodynamical
equilibrium, which corresponds to the opposite limit of vanishing mean free path. All transport coefficients appearing
in the equations of motion are proportional to the mean free path of particles, which is assumed to be much smaller
than the typical length scale over which fluid-dynamical quantities vary. In order to be consistent, the electric
conductivity must be of the same order as the other transport coefficients 
(in fact, the famous Wiedemann-Franz law provides a unique relationship
between the conductivity and the particle
diffusion constant), and in principle we do not have the freedom to send it to infinity.
In case of a finite $\sigma_E$, we are in turn forced to consider a non-vanishing $E^\mu$. 
Nevertheless, since non-resistive magnetohydrodynamics is a theory which is widely applied to physical systems,
we decided to separate the discussion by first treating the somewhat simpler case $E^\mu =0$ 
(corresponding to a non-resistive fluid), which is subject of the present work, and then embarking on
a treatment of the more complicated case $E^\mu \neq 0$, which will be the focus of a follow-up to this paper.

For $E^\mu =0$, the Faraday tensor (\ref{F_munu}) and its Hodge dual (\ref{F_munu_dual}) simplify to
\begin{eqnarray}
F^{\mu \nu } &\longrightarrow & B^{\mu \nu }=\epsilon^{\mu \nu \alpha \beta }u_{\alpha }B_{\beta }\;,  \label{B_munu} \\
\tilde{F}^{\mu \nu } &\longrightarrow &\tilde{B}^{\mu \nu }=B^{\mu }u^{\nu}-B^{\nu }u^{\mu }\;,
\end{eqnarray}
while Maxwell's equations (\ref{Maxwell_inhom}), (\ref{Maxwell_hom}) reduce with Eq.\ (\ref{J_mu_ext}) to 
\begin{eqnarray}
\epsilon^{\mu \nu \alpha \beta } \left( u_\alpha \partial_{\mu }B_\beta
+ B_\beta \partial_\mu u_\alpha \right) &=&\textswab{J}^{\nu }_{ext} + \textswab{J}^\nu_f\;,  \label{Maxwell_inhom_B} \\
\dot{B}^{\mu} + B^\mu \theta &=&u^\mu \partial _{\nu }B^{\nu } + B^\nu \nabla_\nu u^{\mu }\; , \label{Maxwell_hom_B}
\end{eqnarray}
where $\dot{A} \equiv u^\mu \partial_\mu A$ is the comoving derivative of any quantity $A$ and
$\theta \equiv \partial_\mu u^\mu$ is the expansion scalar.

The energy-momentum tensor of the electromagnetic field becomes 
\begin{eqnarray}
T_{em}^{\mu \nu} &\longrightarrow & T_{B}^{\mu \nu } =
\frac{B^{2}}{2}\left( u^{\mu }u^{\nu }-\Delta ^{\mu \nu }-2b^{\mu }b^{\nu}\right)\; ,  \label{T_munu_B}
\end{eqnarray}
where we introduced $B^2 \equiv - B^\mu B_\mu$ and
\begin{equation}
b^{\mu }\equiv \frac{B^{\mu }}{B}\;,  \label{b_mu}
\end{equation}
which is orthogonal to $u^\mu$, $b^\mu u_\mu =0$, and normalized to $b^{\mu }b_{\mu }=-1$.

For systems with a spatial anisotropy, as for instance induced by a magnetic field
\cite{Huang:2011dc,Hernandez:2017mch,Gedalin_1991,Gedalin_1995}, but not necessarily restricted to this 
case \cite{Molnar:2016vvu}
[for a review see Ref.\ \cite{Alqahtani:2017mhy} and refs.\ therein], 
it is convenient to introduce a rank-two operator projecting onto the two-dimensional subspace orthogonal to 
both $u^{\mu }$ and $b^{\mu }$,
\begin{equation}
\Xi^{\mu \nu }\equiv g^{\mu \nu }-u^{\mu }u^{\nu }+b^{\mu }b^{\nu }=\Delta^{\mu \nu }+b^{\mu }b^{\nu }\;.  \label{Xi_munu}
\end{equation}
Furthermore, since $B^{\mu \nu }B_{\mu \nu }=2B^{2}$ it makes sense to
introduce a new dimensionless antisymmetric tensor
\begin{equation}
b^{\mu \nu }\equiv -\frac{B^{\mu \nu }}{B} = - \epsilon^{\mu \nu \alpha \beta} u_\alpha b_\beta \;.  \label{b_munu}
\end{equation}
Obviously, $b^{\mu \nu }u_{\nu }= b^{\mu \nu}b_{\nu }=0$, while Eq.\ (\ref{aux2}) yields
$b^{\mu \nu }b_{\mu \nu }\equiv -2b^{\mu }b_{\mu }=2$. Moreover, with the help of Eq.\ (\ref{aux1}) one can show that
\begin{equation}
b^{\mu \alpha }b_{\nu \alpha }=\Xi _{\nu }^{\mu }\;.
\end{equation}

\subsection{Consequences for energy and momentum evolution of the fluid}

Already at this point we can draw conclusions from the assumption of non-resistivity for the
equations of motion of magnetohydrodynamics. Projecting Eqs.\ (\ref{d_mu_T_munu_em2}), (\ref{d_mu_T_munu_f}) 
onto the direction of $u_\nu$ leads to 
\begin{eqnarray}
\label{u_muT_munuB}
u_{\nu }\partial _{\mu }T_{B}^{\mu \nu } = B\, u_\nu b^{\nu \lambda } (\textswab{J}_{ext, \lambda}
+ \textswab{J}_{f, \lambda})=0\;, \\
\label{u_muT_munuf}
u_{\nu }\partial _{\mu }T_{f}^{\mu \nu } = -B\, u_\nu b^{\nu \lambda }\textswab{J}_{f,\lambda}=0\;, 
\end{eqnarray}
because of $u_\nu b^{\nu \lambda}=0$.
The latter equation means that a magnetic field does not change the fluid energy, which is therefore 
separately conserved.
This is easily understood since a magnetic field (contrary to an electric field) only changes the
direction of the momenta of the particle, but not their energy.
On the other hand, projecting Eqs.\ (\ref{d_mu_T_munu_em2}), (\ref{d_mu_T_munu_f}) onto the three-space 
orthogonal to $u_\nu$ we have
\begin{eqnarray}
\Delta_{\nu }^{\alpha }\partial _{\mu }T_{B}^{\mu \nu } 
=  \left[ B^{2}\dot{u}^{\alpha}-\nabla^{\alpha }\left( \frac{B^{2}}{2}\right) 
-\Delta _{\nu }^{\alpha }\partial_{\mu }\left( B^{2}b^{\mu }b^{\nu }\right) \right] & = &
B\,b^{\alpha \lambda }( \textswab{V}_{ext, \lambda } + \textswab{V}_{f, \lambda})
\; , \label{projDeltad_muT_munuB} \\
\Delta_{\nu }^{\alpha }\partial _{\mu }T_{f}^{\mu \nu }& = & 
-B\, b^{\alpha \lambda} \textswab{V}_{f,\lambda} \label{projDeltad_muT_munuf}\; ,
\end{eqnarray}
where we employed Eq.\ (\ref{d_mu_T_munu_em2}) 
with Eq.\ (\ref{T_munu_B}) to obtain the first equation. For both equations
we used the decomposition (\ref{J_mu_em}), see also Eq.\ (\ref{J_f_mu}), and employed the 
orthogonality $b^{\alpha \lambda} u_\lambda = 0$.

The interpretation of Eq.\ (\ref{projDeltad_muT_munuf})
is that the momentum of the fluid changes on account of the interaction of the magnetic field with
the charge diffusion current. Note that the magnetic field influences the dynamics of the fluid
only by coupling to the \emph{dissipative} part of the charge current. 
Without dissipation, the dynamics of the fluid is unaffected by the magnetic field, 
see Eq.\ (\ref{decoupling}) below.

\subsection{Equations of motion of non-resistive, non-dissipative magnetohydrodynamics}

The equations of motion of non-resistive, non-dissipative magnetohydrodynamics are obtained under the 
assumption that the fluid
is in local thermodynamical equilibrium everywhere in space-time. In the case of dilute gases this
assumption implies that the single-particle distribution function assumes the form \cite{Juttner}
\begin{equation}\label{f_0k}
f_{\mathbf{k}}\longrightarrow f_{0\mathbf{k}}=\left[ \exp \left( \beta _{0}E_{\mathbf{k}}-\alpha _{0}\right) +a\right] ^{-1}\;,  
\end{equation}
with $\alpha _{0}=\mu \beta _{0}$, where $\mu$ is the chemical potential
associated with the particle density $n_{0}$, and $a=\pm 1$ for fermions/bosons, while $a\rightarrow 0$ for classical
particles. Since we assumed that we can neglect the Landau quantization of single-particle energy eigenstates
(see Introduction),
the distribution function is isotropic in the local frame, $E_{\mathbf{k},LR} = \sqrt{\mathbf{k}^2 + m_0^2}$.
Local equilibrium means that the quantities $\alpha_0$, $\beta_0$, as well as the fluid 
velocity $u^\mu$ are functions of the space-time variable $x^\mu$. Since $f_{0\mathbf{k}}$ depends solely on these
five independent variables, and since $N_f^\mu$ and $T^{\mu\nu}_f$ computed from Eqs.\ (\ref{kinetic:N_mu}),
(\ref{kinetic:T_munu}) with $f_{0\mathbf{k}}$
replacing $f_{\mathbf{k}}$ then also depend only on these five variables, the equations of motion of 
magnetohydrodynamics are closed.

In the following, we need the thermodynamic integrals
\begin{equation}
I_{nq}\left( \alpha _{0},\beta _{0}\right) =\frac{\left( -1\right) ^{q}}{\left( 2q+1\right) !!}\left\langle 
E_{\mathbf{k}}^{n-2q}\left( \Delta^{\alpha \beta }k_{\alpha }k_{\beta }\right) ^{q}\right\rangle _{0}\;,
\label{I_nq}
\end{equation}
where $\left\langle \cdots \right\rangle_0 \equiv \int dK\, \cdots\, f_{0\mathbf{k}}$
is defined in analogy to Eq.\ (\ref{average}).
Similarly, the auxiliary thermodynamic integrals are
\begin{equation}
J_{nq}\equiv \left( \frac{\partial I_{nq}}{\partial \alpha _{0}}\right)_{\beta _{0}}
=\frac{\left( -1\right) ^{q}}{\left( 2q+1\right) !!}\left\langle E_{\mathbf{k}}^{n-2q}
\left( \Delta ^{\alpha \beta }k_{\alpha }k_{\beta}\right) ^{q}\left( 1-af_{0\mathbf{k}}\right) \right\rangle_0\; .  \label{J_nq}
\end{equation}
Since $\left( \frac{\partial I_{nq}}{\partial \beta _{0}}\right)_{\alpha _{0}}=-J_{n+1,q}$, 
the total derivative is 
\begin{equation}
dI_{nq}(\alpha_{0},\beta _{0})\equiv \frac{\partial I_{nq}}{\partial \alpha _{0}}\,d\alpha
_{0}+\frac{\partial I_{nq}}{\partial \beta _{0}}\,d\beta
_{0}=J_{nq}\,d\alpha _{0}-J_{n+1,q}\,d\beta _{0}\;.
\end{equation}

Using the equilibrium distribution function in Eqs.\ (\ref{kinetic:N_mu}), (\ref{kinetic:T_munu}) we obtain the 
conserved quantities in the form for a non-dissipative fluid,
\begin{eqnarray}
N_{f0}^{\mu } &\equiv &\left\langle k^{\mu }\right\rangle _{0}=n_{f0}u^{\mu }\;,
\label{kinetic:N_0_mu} \\
T_{f0}^{\mu \nu } &\equiv &\left\langle k^{\mu }k^{\nu }\right\rangle_{0}
=\varepsilon_{0}u^{\mu }u^{\nu }-P_{0}\Delta ^{\mu \nu }\;,  \label{kinetic:T_0_munu}
\end{eqnarray}
where 
\begin{eqnarray}
n_{f0} &\equiv &N_{f0}^{\mu }u_{\mu }=I_{10}\;,
\label{n_0_equilibrium} \\
\varepsilon_{0} &\equiv &T_{f0}^{\mu \nu }u_{\mu }u_{\nu }=I_{20}\;,  \label{e_0_equilibrium} \\
P_{0} &\equiv &-\frac{1}{3}\,T_{f0}^{\mu \nu }\Delta _{\mu \nu}=I_{21} \;.  \label{P_0_equilibrium}
\end{eqnarray}
Therefore, the total energy-momentum tensor of a non-resistive, non-dissipative fluid reads
\begin{equation}
T_{f0+B}^{\mu \nu }\equiv T_{f0}^{\mu \nu }+T_{B}^{\mu \nu }=\left( \varepsilon_{0}+\frac{B^{2}}{2}\right) 
u^{\mu }u^{\nu }-\left( P_{0}+\frac{B^{2}}{2}\right) \Delta ^{\mu \nu }-B^{2}b^{\mu }b^{\nu }\;.
\end{equation}

An immediate consequence of the assumptions of non-resistivity as well as 
non-dissipativity is that the 
energy and momentum of the fluid is separately conserved,
\begin{equation} \label{decoupling}
\partial _{\mu }T_{f0}^{\mu \nu }=0\;.
\end{equation}
This follows immediately from Eq.\ (\ref{d_mu_T_munu_f}), since
$F^{\nu \lambda} \textswab{J}_\lambda \longrightarrow -B \, b^{\nu \lambda} \textswab{n} u_\lambda =0$,
but it also follows from Eqs.\ (\ref{u_muT_munuf}) and (\ref{projDeltad_muT_munuf}), 
since $\textswab{V}_{f,\lambda} \equiv 0$ for a non-dissipative fluid.
The energy of the magnetic field is conserved on account of Eq.\ (\ref{u_muT_munuB}), but the momentum
only when $\textswab{V}_{ext, \lambda} =0$, cf.\ Eq.\ (\ref{projDeltad_muT_munuB}). 

\section{Non-resistive, dissipative magnetohydrodynamics}
\label{Boltzmann_MHD}

In this section, we derive the equations of motion of non-resistive, dissipative magnetohydrodynamics for
a fluid consisting of a single type of point-like particles without dipole moment or spin.
We also assume that the particles undergo binary elastic collisions only. Starting from the Boltzmann equation
in the presence of an external electromagnetic field, we first derive the (infinite)
set of equations of motion for the irreducible moments of the deviation 
\begin{equation} \label{kinetic:f=f0+df}
\delta f_{\mathbf{k}} \equiv f_{\mathbf{k}} - f_{0\mathbf{k}}
\end{equation}
of the single-particle distribution function from isotropic local thermodynamical equilibrium. 
Then we truncate this set using the
14-moment approximation. Our treatment follows closely that of Refs.\ \cite{Denicol:2012cn,Denicol:2012es},
extending the latter by terms arising from the magnetic field. 
Note that our assumption $\beta_0 \ll R_T$ (see Introduction) allows us to neglect Landau quantization,
otherwise $f_{0\mathbf{k}}$ would be anisotropic. 
In principle, however, this case can be discussed using the formalism presented in Ref.\ \cite{Molnar:2016vvu}.
An anisotropy also emerges when using an $f_{0\mathbf{k}}$ which is a solution of the 
Vlasov equation \cite{Gedalin_1991,Gedalin_1995},
or an anisotropic distribution function parametrizing deviations from local equilibrium \cite{Romatschke:2003ms}.

\subsection{Equations of motion for the irreducible moments}
\label{sec:diss_eom_moments}

The relativistic Boltzmann equation coupled to an electromagnetic field \cite{deGroot,Cercignani_book} is
\begin{equation}
k^{\mu }\partial_{\mu }f_{\mathbf{k}}+\textswab{q} F^{\mu \nu }k_{\nu }\frac{\partial }{\partial k^{\mu }}f_{\mathbf{k}}
=C\left[ f\right]\; .
\label{BTE_Fmunu}
\end{equation}
Here the assumption is that the electromagnetic field $F^{\mu \nu}$ changes the momenta $k^\mu$ of particles
carrying charge $\textswab{q}$ on large space-time scales $\sim R_T$, while the collision term, being a quantity 
which is local
in space-time, redistributes them on small space-time scales $\sim \sqrt{\sigma/\pi}$.
We remark that if the particles carry a dipole moment or spin, there would be an additional term on
the left-hand side \cite{Israel:1978up}. Note that for Eq.\ (\ref{BTE_Fmunu}) it does not matter whether
the electromagnetic field is generated exclusively via the charge current of the particles, $\textswab{J}_{f}^\nu$ as
source term in the inhomogeneous Maxwell equations, or exclusively via an external charge current
$\textswab{J}_{ext}^\nu$, or by a combination of both. However, on account of our remarks made at the end of 
Sec.\ \ref{sec:IIB}, only the
case of a non-vanishing external charge current allows to treat the particle current $N_f^\mu$ as an 
independent fluid-dynamical variable.

Under the assumption that the particles undergo binary elastic collisions only, the collision term reads
\begin{equation}
C\left[ f\right] =\frac{1}{2}\int dK^{\prime }dPdP^{\prime }\left[ W_{\mathbf{pp}^{\prime }\rightarrow \mathbf{kk}^{\prime }}
f_{\mathbf{p}}f_{\mathbf{p}^{\prime }}\left( 1-af_{\mathbf{k}}\right) \left( 1-af_{\mathbf{k}^{\prime }}\right) 
-W_{\mathbf{kk}^{\prime }\rightarrow \mathbf{pp}^{\prime}}f_{\mathbf{k}}f_{\mathbf{k}^{\prime }}\left( 1-af_{\mathbf{p}}\right)
\left( 1-af_{\mathbf{p}^{\prime }}\right) \right]\; ,  \label{COLL_INT}
\end{equation}
where the factors $1-af$ represent the corrections from quantum statistics.
The invariant transition rate $W_{\mathbf{kk}^{\prime}\rightarrow \mathbf{pp}^{\prime }}$ satisfies detailed balance, 
$W_{\mathbf{kk}^{\prime }\rightarrow \mathbf{pp}^{\prime }}=W_{\mathbf{pp}^{\prime}\rightarrow \mathbf{kk}^{\prime }}$, 
and is symmetric with respect to the exchange of momenta, $W_{\mathbf{kk}^{\prime }\rightarrow \mathbf{pp}^{\prime }}
=W_{\mathbf{k}^{\prime }\mathbf{k}\rightarrow \mathbf{pp}^{\prime}}
=W_{\mathbf{kk}^{\prime }\rightarrow \mathbf{p}^{\prime }\mathbf{p}}$. 

Following Refs.\ \cite{Denicol:2012cn,Denicol:2012es} 
we define the irreducible moments of $\delta f_{\mathbf{k}}$ as~\footnote{A tensor is called irreducible 
when it is irreducible under a group $G$ consisting of Lorentz transformations that leave $u^\mu$ invariant. 
Let $F$ be a subgroup of $G$ consisting of Lorentz transformations that leave both $u^\mu$ and $b^\mu$ invariant. 
An irreducible tensor under $G$ may be reducible under $F$. This reduction of symmetry
leads to a larger number of 
transport coefficients in dissipative magnetohydrodynamics 
than in ordinary dissipative fluid dynamics; see Sec.~\ref{sec:diss_eom_NS}. }
\begin{eqnarray}
\rho _{r}^{\mu _{1}\cdots \mu _{n} } &\equiv
& \left\langle E_{\mathbf{k}}^{r}k^{\left\langle \mu_{1}\right. }\cdots k^{\left. \mu _{n}\right\rangle }\right\rangle _{\delta }\;,
\label{rho_i_irr_moments}
\end{eqnarray}
where $\left\langle \cdots \right\rangle _{\delta }=\int dK\cdots \delta f_{\mathbf{k}}$.
Here, the irreducible tensor of rank $\ell$ is defined as 
\begin{equation}
k^{\left\langle \mu _{1}\right. }\cdots k^{\left. \mu _{\ell }\right\rangle}
=\Delta _{\nu _{1}\cdots \nu _{\ell }}^{\mu _{1}\cdots \mu _{\ell }}k^{\nu_{1}}\cdots k^{\nu _{\ell }}\;,  \label{iso_irreducible_momenta}
\end{equation}
where the rank-$2 \ell$ symmetric and traceless projection tensor $\Delta _{\nu _{1}\cdots \nu _{\ell }}^{\mu _{1}\cdots
\mu _{\ell }}$ is a straightforward generalization of the rank-four projection tensor $\Delta^{\mu \nu}_{\alpha \beta}$
introduced above [for more details on how to construct the former, see Refs.\ \cite{deGroot,Molnar:2016vvu}].
The irreducible tensors $1,\,k^{\left\langle \mu \right\rangle },\,k^{\left\langle \mu \right.}k^{\left. \nu \right\rangle },
\,k^{\left\langle \mu \right. }k^{\nu}k^{\left. \lambda \right\rangle },\ldots $ form a complete basis in momentum space 
and satisfy the following orthogonality condition
\begin{equation}
\int dK\text{ }\mathrm{F}(E_{\mathbf{k}})\ k^{\left\langle \mu _{1}\right.}\cdots 
k^{\left. \mu _{\ell }\right\rangle }k_{\left\langle \nu _{1}\right.}\cdots k_{\left. \nu _{n}\right\rangle }
=\frac{\ell !\ \delta _{\ell n}}{\left( 2\ell +1\right) !!}
\Delta _{\nu _{1}\cdots \nu _{\ell }}^{\mu_{1}\cdots \mu _{\ell }}\int dK\ \mathrm{F}(E_{\mathbf{k}})\ 
\left( \Delta^{\alpha \beta }k_{\alpha }k_{\beta }\right) ^{\ell }\;,
\label{normalization_isotropic}
\end{equation}
where $\mathrm{F}(E_{\mathbf{k}})$ is a sufficiently rapidly converging (but otherwise arbitrary) function
of $E_{\mathbf{k}}$.

The deviations of the particle four-current and the fluid energy-momentum tensor from their
local equilibrium values $N_{f0}^\mu$, $T_{f0}^{\mu \nu}$ are
\begin{eqnarray}
\delta N_{f}^{\mu } &\equiv &\left\langle k^{\mu }\right\rangle _{\delta}
=\delta n_f\, u^{\mu }+V_f^{\mu }\; ,  \label{kinetic:deltaN_mu} \\
\delta T_{f}^{\mu \nu } &\equiv &\left\langle k^{\mu }k^{\nu }\right\rangle_{\delta }
=\delta \varepsilon\, u^{\mu }u^{\nu }-\Pi \,\Delta ^{\mu \nu }+W^{\mu} u^{\nu} +W^{\nu}u^{\mu}+\pi ^{\mu \nu }\; ,
\label{kinetic:deltaT_munu}
\end{eqnarray}
where the corrections to particle density, energy density, and isotropic pressure are 
\begin{eqnarray}
\delta n_f &\equiv &\delta N_{f}^{\mu }u_{\mu } = \rho _{1}\; ,
\label{kinetic:rho_1} \\
\delta \varepsilon &\equiv &\delta T_{f}^{\mu \nu }u_{\mu }u_{\nu }=\rho _{2}\; ,
\label{kinetic:rho_2} \\
\Pi &\equiv &-\frac{1}{3}\delta T_{f}^{\mu \nu }\Delta _{\mu \nu } =-\frac{m_{0}^{2}}{3}\rho _{0}
+ \frac{\rho_2}{3}\;.  \label{kinetic:rho_0}
\end{eqnarray}
The particle and energy-momentum diffusion currents orthogonal to the fluid velocity are 
\begin{eqnarray}
V_f^{\mu } &\equiv &\Delta _{\nu }^{\mu }\delta N_{f}^{\nu }=\rho_{0}^{\mu}\; ,  \label{kinetic:rho_0_mu} \\
W^{\mu } &\equiv &\Delta _{\alpha }^{\mu }\delta T_{f}^{\alpha \beta}u_{\beta }
= \rho _{1}^{\mu}\; ,
\label{kinetic:rho_1_mu}
\end{eqnarray}
while the shear-stress tensor is 
\begin{equation}
\pi^{\mu \nu }\equiv \Delta _{\alpha \beta }^{\mu \nu }\delta T_{f}^{\alpha\beta }=\rho _{0}^{\mu \nu }\;.
\label{kinetic:rho_0_munu}
\end{equation}

Choosing the Landau frame \cite{Landau_book} to determine the fluid velocity implies
\begin{equation}
u^{\mu }=\frac{T_{f}^{\mu \nu }u_{\nu }}{\sqrt{u_{\alpha }T_{f}^{\alpha
\beta }T_{f,\beta \gamma }u^{\gamma }}}\;,\;\;\;\;\rho _{1}^{\mu}=0\; .  \label{LR_Landau}
\end{equation}
The parameters $\alpha_0$ and $\beta_0$ entering $f_{0\mathbf{k}}$ are determined by the so-called
Landau matching conditions, i.e., demanding that the particle density
and energy density resulting from $f_{\mathbf{k}}$ are identical with those resulting from $f_{0\mathbf{k}}$,
$n_f=n_{f0},\, \varepsilon = \varepsilon_0$, or in other words
\begin{equation}
\delta n_f = \rho_{1}=0\; ,\;\;\;\;\ \delta \varepsilon = \rho _{2}=0\;.  \label{Landau_matching}
\end{equation}

Then, the charge four-current and total energy-momentum
tensor in non-resistive, dissipative magnetohydrodynamics are
\begin{eqnarray}
\textswab{J}_f^\mu & \equiv & \textswab{n}_f u^\mu + \textswab{V}_f^\mu\; , \label{J_mu_MHD2} \\
T^{\mu \nu } &\equiv &T_{f}^{\mu \nu }+T_{B}^{\mu \nu } 
=\left( \varepsilon_{0}+\frac{B^{2}}{2}\right) u^{\mu }u^{\nu }-\left( P_{0}+\Pi +\frac{B^{2}}{2}\right) \Delta ^{\mu \nu }
-B^{2}b^{\mu }b^{\nu }+\pi ^{\mu\nu }\;.  \label{T_munu_MHD}
\end{eqnarray}
Equations (\ref{J_f_mucons}), (\ref{d_mu_T_munu_f}) with Eqs.\ (\ref{J_mu_MHD2}), (\ref{T_munu_MHD}) 
together with the thermodynamical 
identities (\ref{I_nq}), (\ref{J_nq}) lead to the following equations of motion for
$\alpha_0$, $\beta_0$, and $u^\mu$:
\begin{eqnarray}
\dot{\alpha}_{0} &=&\frac{1}{D_{20}}\left[ -J_{30}\left( n_{f0}\theta
+\partial _{\mu }V_f^{\mu }\right) +J_{20}\left( \varepsilon_{0}+P_{0}+\Pi \right)
\theta -J_{20}\pi ^{\mu \nu }\sigma _{\mu \nu }\right] \; ,  \label{D_alpha} \\
\dot{\beta}_{0} &=&\frac{1}{D_{20}}\left[ -J_{20}\left( n_{f0}\theta
+\partial _{\mu }V_f^{\mu }\right) +J_{10}\left( \varepsilon_{0}+P_{0}+\Pi \right)
\theta -J_{10}\pi ^{\mu \nu }\sigma _{\mu \nu }\right]\; ,  \label{D_beta}
\end{eqnarray}
and 
\begin{equation}
\dot{u}^{\mu }=\frac{1}{\varepsilon_0+P_0} \left[ \frac{n_{f0}}{\beta _{0}}\left( \nabla ^{\mu }\alpha_{0}- 
h_0 \nabla ^{\mu }\beta _{0}\right) 
- \Delta_{\nu }^{\mu }\partial _{\kappa }\pi ^{\kappa \nu }-\Pi \dot{u}^{\mu}+\nabla ^{\mu }\Pi 
- \textswab{q}B\, b^{\mu \nu}V_{f, \nu }\right] \;,  \label{D_u_mu}
\end{equation}
where $D_{nq} \equiv J_{n+1,q}J_{n-1,q} - J_{nq}^2$, 
$h_{0}\equiv \left( \varepsilon_{0}+P_{0}\right) /n_{f0}$ is the enthalpy per particle, and 
$\sigma ^{\mu \nu }=\nabla ^{\left\langle \mu \right.}u^{\left. \nu \right\rangle }$ is the shear tensor. 
The equations of motion for  $\alpha_{0}$ and $\beta_{0}$ 
are the same as Eqs.\ (39), (40) of Ref.\ \cite{Denicol:2012cn}, however Eq.\ (\ref{D_u_mu}) contains
an additional term due to the magnetic field when compared to Eq.\ (41) of Ref.\ \cite{Denicol:2012cn}.

We now use Eq.\ (\ref{kinetic:f=f0+df}) to replace $f_{\mathbf{k}}$
by $\delta f_{\mathbf{k}}$ in the Boltzmann equation (\ref{BTE_Fmunu}).
Then, we take moments of the Boltzmann equation (\ref{BTE_Fmunu}) in momentum space.
With the definitions 
\begin{equation}
\dot{\rho}_{r}^{\left\langle \mu _{1}\cdots \mu _{\ell}\right\rangle }
\equiv \Delta _{\nu_{1}\cdots \nu _{\ell }}^{\mu _{1}\cdots \mu _{\ell }} u^\alpha \partial_\alpha 
\rho _{r}^{\nu_{1}\cdots \nu _{\ell}}\; ,  
\label{Dtau_rho_general}
\end{equation}
and
\begin{equation}
\mathcal{C}_{r}^{\left\langle \mu _{1}\cdots \mu _{\ell }\right\rangle}\equiv \Delta _{\nu _{1}\cdots \nu _{\ell }}^{\mu _{1}
\cdots \mu _{\ell }}\int dK\,E_{\mathbf{k}}^{r}\,k^{\nu _{1}}\cdots k^{\nu _{\ell }}C\left[ f\right]\; ,  \label{Irr_coll_int}
\end{equation}
we obtain the equations of motion for the irreducible moments, similarly as shown in 
Refs.\ \cite{Denicol:2012cn,Denicol:2012es}.

The equation of motion for the irreducible tensors of rank zero reads
\begin{align}
\dot{\rho}_{r}-C_{r-1}& =\alpha _{r}^{\left( 0\right) }\theta +\frac{G_{3r}}{D_{20}}
\partial _{\mu }V_f^{\mu }+\frac{\theta }{3}\left[ m_0^{2}(r-1)\rho_{r-2}-(r+2)\rho _{r}-3\frac{G_{2r}}{D_{20}}\Pi \right]  \notag \\
& +r\rho _{r-1}^{\mu }\dot{u}_{\mu }-\nabla _{\mu }\rho _{r-1}^{\mu }
+\left[(r-1)\rho _{r-2}^{\mu \nu }+\frac{G_{2r}}{D_{20}}\pi ^{\mu \nu }\right]
\sigma _{\mu \nu }\;.  \label{D_rho}
\end{align}
where we have defined $G_{nm} =J_{n,0}J_{m,0}-J_{n-1,0}J_{m+1,0}$.
Note that the contribution of the magnetic field
vanishes for any scalar moment and exactly corresponds to Eq.\ (35) of Ref.\ \cite{Denicol:2012cn}. 
However, the magnetic field is still present and
affects the fluid motion through the acceleration equation, Eq.\ (\ref{D_u_mu}), as well as through the equations of
motion for the irreducible moments of rank higher than zero, see below.

The equation of motion for the irreducible tensors of rank one is
\begin{align}
\dot{\rho}_{r}^{\left\langle \mu \right\rangle }-C_{r-1}^{\left\langle \mu\right\rangle }& 
=\alpha _{r}^{\left( 1\right) }\nabla ^{\mu }\alpha_{0}+r\rho _{r-1}^{\mu \nu }\dot{u}_{\nu }
-\frac{1}{3}\nabla ^{\mu }\left[m_0^{2}\rho _{r-1}-\rho _{r+1}\right] -\Delta _{\alpha }^{\mu }
\left( \nabla_{\nu }\rho _{r-1}^{\alpha \nu }+\alpha _{r}^{h}\partial _{\kappa }\pi^{\kappa \alpha }\right)  \notag \\
& +\frac{1}{3}\left[ m_{0}^{2}\left( r-1\right) \rho _{r-2}^{\mu }-\left(r+3\right) \rho _{r}^{\mu }\right] \theta 
+\left( r-1\right) \rho_{r-2}^{\mu \nu \lambda }\sigma _{\mu \nu }  \notag \\
& +\frac{1}{5}\sigma ^{\mu \nu }\left[ m_{0}^{2}\left( 2r-2\right) \rho_{r-2,\nu }
-\left( 2r+3\right) \rho _{r,\nu }\right] +\rho _{r,\nu }\omega^{\mu \nu }  \notag \\
& +\frac{1}{3}\left[ m_{0}^{2}r\rho _{r-1}-\left( r+3\right) \rho_{r+1}-3\alpha _{r}^{h}\Pi \right] \dot{u}^{\mu }
+\alpha _{r}^{h}\nabla^{\mu }\Pi  \notag \\
& -\alpha _{r}^{h}\, \textswab{q}Bb^{\mu \nu }V_{f, \nu }-\textswab{q}Bb^{\mu \nu }\rho_{r-1,\nu }\;,  \label{D_rho_mu}
\end{align}
where $\omega ^{\mu \nu }=(\nabla^{ \mu  }u^{ \nu } - \nabla^\nu u^\mu)/2$ is the vorticity tensor. 
The two terms in the last line are new as compared to Eq.\ (36) of Ref.\ \cite{Denicol:2012cn}
and  explicitly contain the magnetic field.

The equation of motion for the irreducible moments of tensor of rank two is
\begin{align}
\dot{\rho}_{r}^{\left\langle \mu \nu \right\rangle }-C_{r-1}^{\left\langle\mu \nu \right\rangle }& 
=2\alpha _{r}^{\left( 2\right) }\sigma ^{\mu \nu }+\frac{2}{15}\left[ m_{0}^{4}\left( r-1\right) 
\rho _{r-2}-\left( 2r+3\right)m_{0}^{2}\rho _{r}+\left( r+4\right) \rho _{r+2}\right] \sigma ^{\mu \nu } 
\notag \\
& +\frac{2}{5}\dot{u}^{\left\langle \mu \right. }\left[ m_{0}^{2}r\rho_{r-1}^{\left. \nu \right\rangle }
-\left( r+5\right) \rho _{r+1}^{\left. \nu\right\rangle }\right] 
-\frac{2}{5}\left[ \nabla ^{\left\langle \mu \right.}\left( m_{0}^{2}\rho _{r-1}^{\left. \nu \right\rangle }
-\rho _{r+1}^{\left.\nu \right\rangle }\right) \right]  \notag \\
& +r\rho _{r-1}^{\mu \nu \gamma }\dot{u}_{\gamma }-\Delta _{\alpha \beta}^{\mu \nu }
\nabla _{\lambda }\rho _{r-1}^{\alpha \beta \lambda }
+\left(r-1\right) \rho _{r-2}^{\mu \nu \lambda \kappa }\sigma _{\lambda \kappa}
+2\rho _{r}^{\lambda \left\langle \mu \right. }
\omega _{\hspace*{0.2cm}\lambda }^{\left. \nu \right\rangle }  \notag \\
& +\frac{1}{3}\left[ m_{0}^{2}\left( r-1\right) \rho _{r-2}^{\mu \nu}
-\left( r+4\right) \rho _{r}^{\mu \nu }\right] \theta 
+\frac{2}{7}\left[m_{0}^{2}\left( 2r-2\right) \rho _{r-2}^{\kappa \left\langle \mu \right.}
-\left( 2r+5\right) \rho _{r}^{\kappa \left\langle \mu \right. }\right]\sigma _{\kappa }^{\left. \nu \right\rangle }  \notag \\
& -2\,\textswab{q}Bb^{\alpha \beta }\Delta _{\alpha \kappa }^{\mu \nu }g_{\lambda\beta }\rho _{r-1}^{\kappa \lambda }\;,
 \label{D_rho_munu}
\end{align}
where only the last term is new when compared to Eq.\ (37) of Ref.\ \cite{Denicol:2012cn} 
and explicitly contains the magnetic field.
Here we also defined the following coefficients which are formally unchanged
from Eqs.\ (42) -- (44) of Ref.\ \cite{Denicol:2012cn}, 
\begin{align}
\alpha _{r}^{\left( 0\right) }& =\left( 1-r\right) I_{r1}-I_{r0}-\frac{n_{f0}}{D_{20}}\left( h_{0}G_{2r}-G_{3r}\right)\; ,  
\label{alpha_i_0} \\
\alpha _{r}^{\left( 1\right) }& =J_{r+1,1}-h_{0}^{-1}J_{r+2,1}\; ,
\label{alpha_i_1} \\
\alpha _{r}^{\left( 2\right) }& =I_{r+2,1}+\left( r-1\right) I_{r+2,2}\; ,
\label{alpha_i_2} \\
\alpha _{r}^{h}& =-\frac{\beta _{0}}{\varepsilon_{0}+P_{0}}J_{r+2,1}\;.
\label{alpha_i_h}
\end{align}
The collision integral can be linearized using Eq.\ (\ref{kinetic:f=f0+df}) and written as
\begin{equation}
C_{r-1}^{\left\langle \mu _{1}\cdots \mu _{\ell }\right\rangle }\equiv
-\sum_{n=0}^{N_{\ell }}\mathcal{A}_{rn}^{\left( \ell \right) }\rho _{n}^{\mu_{1}\cdots \mu _{\ell }}\; ,  \label{Lin_collint}
\end{equation}
where the coefficient $\mathcal{A}_{rn}^{\left( \ell \right) }$ contains time scales $\sim \lambda_{\rm mfp}$.
In order to obtain this result, we have assumed that the magnetic field does not modify the
collision integral, so that we were able to employ the orthogonality relation 
(\ref{normalization_isotropic}), for details see Ref.\ \cite{Denicol:2012cn}. 

%Since the fluid-dynamical equations of motion do not involve tensors of rank higher than two, we also restrict
%ourselves to the equations of motion (\ref{D_rho}) -- (\ref{D_rho_munu}). Higher-rank
%tensors on the right-hand sides of these equations will subsequently be neglected.

Note that, once the equations of motion (\ref{D_rho}) -- (\ref{D_rho_munu}) (and in principle those for
all higher-rank tensors) are solved and the complete set of irreducible moments is determined, one can 
reconstruct the single-particle distribution $f_{\mathbf{k}}$ as a solution of the Boltzmann equation.
Following Refs.\ \cite{Denicol:2012cn,Denicol:2012es},
\begin{equation}
f_{\mathbf{k}}= f_{0\mathbf{k}}+f_{0\mathbf{k}}\left( 1-af_{0\mathbf{k}}\right) 
\sum_{\ell =0}^{\infty }\sum_{n=0}^{N_{\ell }}\rho _{n}^{\mu_{1}\cdots \mu _{\ell }}
k_{\left\langle \mu _{1}\right. }\cdots k_{\left.\mu _{\ell }\right\rangle }\mathcal{H}_{\mathbf{k}n}^{(\ell )}\; .
\label{f_iso_expansion}
\end{equation}
We remark that this relation is an exact equality (i.e., $f_{\mathbf{k}}$ an exact solution of the
Boltzmann equation) only if we take $N_\ell \rightarrow \infty$. In practice, however, one has to 
truncate the sum over $n$ at some finite value, $N_{\ell }< \infty$. The same holds for the sum over $\ell$.
Since there are no tensors of rank higher than two in fluid dynamics, this sum is usually restricted to $\ell \leq 2$. 
Furthermore, this also implies that higher-rank tensors on the right-hand sides of the 
equations of motion (\ref{D_rho}) -- (\ref{D_rho_munu}) will be subsequently neglected.

The coefficients $\mathcal{H}_{\mathbf{k}n}^{(\ell )}$ are defined as 
\begin{equation}
\mathcal{H}_{\mathbf{k}n}^{(\ell )}=\frac{\left( -1\right) ^{\ell }}{\ell !\
J_{2\ell ,\ell }}\sum_{i=n}^{N_{\ell }}\sum_{m=0}^{i}a_{in}^{(\ell)}a_{im}^{(\ell )}E_{\mathbf{k}}^{m}\; ,  \label{H_kn}
\end{equation}
where the coefficients $a_{ij}^{(\ell )}$ can be written in terms of
thermodynamic integrals and are calculated via Gram-Schmidt
orthogonalization, for details see Ref.\ \cite{Denicol:2012cn}. 

In preparation of a suitable truncation of the infinite set of equations of motion for the irreducible moments, we note
that an irreducible moment of arbitrary order $r$ and tensor rank $\ell$ can always be expressed as a linear
combination of irreducible moments of all orders $n$ and the same tensor rank,
\begin{equation} \label{useful}
\rho _{r}^{\mu _{1}\cdots \mu _{\ell }} = \sum_{n=0}^{N_{\ell }}\rho_{n}^{\mu _{1}\cdots \mu _{\ell }}
\mathcal{F}_{-r,n}^{\left( \ell \right)}=\sum_{n=0}^{N_{\ell }}\rho _{n}^{\mu _{1}\cdots \mu _{\ell}}\sum_{i=n}^{N_{\ell }}
\sum_{m=0}^{i}a_{in}^{(\ell )}a_{im}^{(\ell )}\frac{J_{r+m+2\ell ,\ell }}{J_{2\ell ,\ell }}\;,
\end{equation}
where 
\begin{equation}
\mathcal{F}_{rn}^{\left( \ell \right) }=\frac{\ell !}{\left( 2\ell +1\right)!!}
\int dK E_{\mathbf{k}}^{-r}\mathcal{H}_{\mathbf{k}n}^{\left( \ell \right)}
\left( \Delta ^{\alpha \beta }k_{\alpha }k_{\beta }\right) ^{\ell }f_{0\mathbf{k}}\left( 1-af_{0\mathbf{k}}\right) \;.
\end{equation}
The first equality of relation (\ref{useful}) is proven using the orthogonality (\ref{normalization_isotropic}) of the 
irreducible moments and their definition (\ref{rho_i_irr_moments}). The second equality of relation
(\ref{useful}) is shown using the definitions of the auxiliary thermodynamic integrals (\ref{J_nq}) and of the 
coefficients (\ref{H_kn}). Note that Eq.\ (\ref{useful}) is an identity for $0 \leq r \leq N_\ell$, while it is an
approximation for $r$ outside this range, unless $N_\ell \rightarrow \infty$. The accuracy of this approximation
can be systematically improved by increasing $N_\ell$.
In the remainder of this paper, however, we will restrict ourselves to the so-called
14-moment approximation, i.e., we will assume $N_0 = 2$, $N_1=1$, and $N_2 =0$ \cite{Denicol:2012cn}.

\subsection{The Navier-Stokes approximation}
\label{sec:diss_eom_NS}

Besides a suitable truncation of Eqs.\ (\ref{D_rho}) -- (\ref{D_rho_munu}), we also need a scheme to power count
the various terms in these equations, in order to define the order of the approximation we are considering. 
We assume that quantities representing deviations from local thermodynamical 
equilibrium, like the irreducible moments, are of first order in some small parameter. 
Furthermore, since macroscopic fields like $\alpha_0(x^\mu),\, \beta_0(x^\mu)$, and $u^\mu(x^\mu)$ 
vary on space-time scales that are much larger than the microscopic scales contained in the collision integral,
we also assume that derivatives of these fields are of first order in that small parameter.

In the Navier-Stokes approximation, all second-order terms, i.e., terms involving products of irreducible moments and
derivatives of $\alpha_0, \, \beta_0$, and $u^\mu$, or derivatives of irreducible moments are
neglected, leaving only the collision integrals [in linearized form, see Eq.\ (\ref{Lin_collint})] on the left-hand sides 
and the first terms as well as the last terms involving the magnetic field 
on the right-hand sides of Eqs.\ (\ref{D_rho}) -- (\ref{D_rho_munu}). Bringing the latter ones to the left-hand side
results in the following set of equations,
\begin{eqnarray}
\sum_{n=0,\neq 1,2}^{N_{0}}\mathcal{A}_{rn}^{\left( 0\right) }\rho _{n}
&=&\alpha _{r}^{\left( 0\right) }\theta \; ,  \label{asy_rho} \\
\sum_{n=0,\neq 1}^{N_{1}}\left[ \mathcal{A}_{rn}^{\left( 1\right) }g^{\mu \nu}
+\textswab{q}B\left( \mathcal{F}_{1-r,n}^{\left( 1\right) }+\alpha_{r}^{h}\delta _{n0}\right) b^{\mu \nu }\right]  \rho _{n,\nu }
&=&\alpha_{r}^{\left( 1\right) }\nabla ^{\mu }\alpha _{0}\; ,  \label{asy_rho_mu} \\
\sum_{n=0}^{N_{2}}\left[ \mathcal{A}_{rn}^{\left( 2\right) } g^\mu_\alpha g^\nu_\beta
+\textswab{q}B\mathcal{F}_{1-r,n}^{\left( 2\right) }\left( b_{\left. {}\right.\beta }^{\mu }g^{\nu}_{\alpha}
+b_{\left. {}\right. \beta}^{\nu}g^{\mu}_{\alpha }\right) \right]  \rho _{n}^{\alpha\beta }
&=&2\alpha _{r}^{\left( 2\right)}\sigma ^{\mu \nu }\;.  \label{asy_rho_mu_nu}
\end{eqnarray}
In physical terms, it is assumed that the irreducible moments no longer evolve in time and assume
their asymptotic solution given solely by the first-order terms on the right-hand side, multiplied by the
inverse of the coefficient matrix on the left-hand side.
The formal solution of this set of equations is
\begin{eqnarray}
\rho _{r} &=&\zeta _{r}^{\mu \nu }\partial _{\mu }u_{\nu }\; ,  \label{rho_NS} \\
\rho _{r}^{\mu } &=&\kappa _{r}^{\mu \nu }\nabla _{\nu }\alpha_0\; , \label{rho_NS_mu} \\
\rho _{r}^{\mu \nu } &=&\eta _{r}^{\mu \nu \alpha \beta }\sigma _{\alpha\beta }\; ,  \label{rho_NS_mu_nu}
\end{eqnarray}
where the rank-two tensor coefficients can in general be decomposed in terms
of the projection operators $\Xi^{\mu \nu}$, $b^\mu b^\nu$, as well as the
tensor $b^{\mu \nu}$ \cite{Huang:2011dc},
\begin{eqnarray}
\zeta_{r}^{\mu \nu }&=& \zeta_{r\perp }\Xi ^{\mu \nu }-\zeta_{r\parallel }b^{\mu}b^{\nu }-\zeta_{r\times }b^{\mu \nu }\;, 
\label{bulkvistensor}\\
\kappa _{r}^{\mu \nu }& =& \kappa _{r\perp }\Xi ^{\mu \nu }-\kappa _{r\parallel}b^{\mu }b^{\nu }
-\kappa _{r\times }b^{\mu \nu }\;, 
\end{eqnarray}
while the rank-4 tensor coefficient involves the projection operator $\Delta^{\mu \nu \alpha \beta}$ and
products of $\Delta^{\mu \nu}$, $\Xi^{\mu \nu}$, $b^\mu b^\nu$, as well as $b^{\mu \nu}$, for more details, see
Ref.\ \cite{Huang:2011dc},
\begin{align}
\eta _{r}^{\mu \nu \alpha \beta }& =2\eta _{r0}\, \Delta^{\mu \nu \alpha \beta}
+\eta _{r1}\left( \Delta ^{\mu \nu }-\frac{3}{2}\Xi ^{\mu \nu }\right)
\left( \Delta ^{\alpha \beta }-\frac{3}{2}\Xi ^{\alpha \beta }\right)  
\notag \\
& -2\eta _{r2}\left( \Xi ^{\mu \alpha }b^{\nu }b^{\beta }+\Xi ^{\nu \alpha
}b^{\mu }b^{\beta }\right) -2\eta _{r3}\left( \Xi ^{\mu \alpha }b^{\nu \beta
}+\Xi ^{\nu \alpha }b^{\mu \beta }\right) +2\eta _{r4}\left( b^{\mu \alpha
}b^{\nu }b^{\beta }+b^{\nu \alpha }b^{\mu }b^{\beta }\right) \; .
\end{align}
The scalar transport coefficients $\zeta_{r\perp },\,  \zeta_{r\parallel }, \, \zeta_{r\times },\, \kappa _{r\perp },\,  
\kappa_{r\parallel }, \, \kappa_{r\times },\,\eta_{r0},\, \eta_{r1},\,\eta_{r2},\,\eta_{r3},\,\eta_{r4}$ 
are obtained by substituting Eqs.\ (\ref{rho_NS}) -- (\ref{rho_NS_mu_nu}) into Eqs.\
(\ref{asy_rho}) -- (\ref{asy_rho_mu_nu}) 
and identifying the coefficients of the corresponding tensor structures. 

The bulk-viscosity coefficients $\zeta_{r\perp },\,  \zeta_{r\parallel }, \, \zeta_{r\times }$ 
are then determined by the following equations,
\begin{align}
\sum_{n=0,\neq 1,2}^{N_{0}}\mathcal{A}_{rn}^{\left( 0\right) }\zeta_{n\perp}& =\alpha _{r}^{\left( 0\right) } \;, \notag \\
\sum_{n=0,\neq 1,2}^{N_{0}}\mathcal{A}_{rn}^{\left( 0\right) }\left( \zeta_{n\perp }-\zeta _{n\parallel }\right) & =0\; ,  \notag \\
\sum_{n=0,\neq 1,2}^{N_{0}}\mathcal{A}_{rn}^{\left( 0\right) }\zeta_{n\times}& =0\;.
\end{align}
and hence in the 14-moment approximation ($N_{0}=2$), 
\begin{equation}
\zeta_{0\perp }  =\zeta_{0\parallel } =\frac{\alpha _{r}^{\left( 0\right) }}{\mathcal{A}_{r0}^{\left( 0\right) }}\;, \;\;\;
\zeta_{0\times } =0\; .
\end{equation}
Note that, as long as the collision integral is assumed to be independent of the magnetic field, only a tensor structure 
of the type $\sim \Xi^{\mu \nu} - b^\mu b^\nu \equiv \Delta^{\mu \nu}$ survives in the bulk-viscosity tensor 
(\ref{bulkvistensor}). In general, however, this does not need to be the case. An explicit example is given in  Ref.~\cite{Hattori:2017qih} 
where $\zeta_{0\perp }$ and 
$\zeta_{0\parallel }$ are calculated for a hot quark-gluon plasma in a magnetic field, taking into account  
Landau quantization.

The transport coefficients $\kappa _{r\perp },\,  \kappa_{r\parallel }, \, \kappa_{r\times }$
are found from Eq.\ (\ref{asy_rho_mu}) by inserting Eq.\ (\ref{rho_NS_mu}). 
This leads to the following system of coupled equations,
\begin{eqnarray}
\sum_{n=0,\neq 1}^{N_{1}}\left[ \mathcal{A}_{rn}^{\left( 1\right) }\kappa_{n\perp }
+\textswab{q}B\left( \mathcal{F}_{1-r,n}^{\left( 1\right) } + \alpha_r^h \delta_{n0}\right) \kappa _{n\times}\right] 
&=&\alpha _{r}^{\left( 1\right) }\;, \\
\sum_{n=0,\neq 1}^{N_{1}} \mathcal{A}_{rn}^{\left( 1\right) }\kappa_{n\parallel }
&=&\alpha_r^{(1)}\; , \\
\sum_{n=0,\neq 1}^{N_{1}}\left[ \mathcal{A}_{rn}^{\left( 1\right) }\kappa_{n\times }
-\textswab{q}B\left( \mathcal{F}_{1-r,n}^{\left( 1\right) } + \alpha_r^h \delta_{n0} \right)\kappa _{n\perp}\right] &=&0\;,
\end{eqnarray}
and hence, in the 14-moment approximation ($N_{1}=1$),
\begin{equation}
\kappa_r \equiv \kappa _{0\parallel } = \frac{\alpha _{r}^{\left( 1\right) }}{\mathcal{A}_{r0}^{\left( 1\right) }}\; , \;\;\;
\kappa _{0\perp } =\kappa _{0\parallel } \left[1+\left( \textswab{q}B
\frac{\mathcal{F}_{1-r,0}^{\left( 1\right) }+\alpha_r^h}{\mathcal{A}_{r0}^{\left( 1\right) }}\right) ^{2}\right]^{-1}\;, \;\;\;
\kappa _{0\times } = \kappa _{0\perp} 
\, \textswab{q}B\frac{\mathcal{F}_{1-r,0}^{\left(1\right) }+\alpha_r^h}{\mathcal{A}_{r0}^{\left( 1\right) }} \; .
\end{equation}
One observes that, when $B\rightarrow 0$, $\kappa_{0\times} \rightarrow 0$, while
$\kappa_{0\parallel} \rightarrow  \kappa_{0\perp}$. Also in this case, the diffusion tensor
$\kappa_0^{\mu \nu} \sim \Delta^{\mu \nu}$, as expected. Moreover, for any $B \neq 0$, 
$\kappa_{0\perp} < \kappa_{0\parallel}$, i.e., due to the cyclotron motion of the particles, 
particle (or charge) diffusion transverse to
the magnetic field is reduced as compared to the diffusion parallel to the magnetic field. 

In the limit of a massless Boltzmann gas, where 
$J_{nq}\equiv I_{nq}=\frac{\left(n+1\right) !}{2 \left( 2q+1\right) !!}\beta_0^{2-n} P_{0}$,  
and for a constant binary cross section $\sigma=const.$, we obtain for $r=0$ the following expressions:
$\alpha_0^{(1)} = \beta_0 P_0 /12$, $\alpha_0^h \equiv -1/h_0 = - \beta_0/4$, 
$\mathcal{F}_{10}^{\left( 1\right) }=2\beta _{0}/3$,
and $\mathcal{A}_{00}^{\left(1\right) }=4/(9\lambda_{\mathrm{mfp}})$, 
where $\lambda_{\mathrm{mfp}}=1/(n_{f0} \sigma)$ is the
mean free path of the particles, and thus the diffusion coefficients assume the values
\begin{equation}
\kappa _{0\parallel } = \frac{3 \lambda_{\mathrm{mfp}}n_{f0} }{16}\;, \;\;\;
\kappa _{0\perp } = \frac{48 \lambda_{\mathrm{mfp}}n_{f0} }{256
+ 225 \xi_B^{2} }\;, \;\;\;
\kappa _{0\times } =\frac{45\, \xi_B \lambda_{\mathrm{mfp}} n_{f0} }{
256+225 \xi_B^{2}}\; ,
\end{equation}
where $\xi_B \equiv \textswab{q}B \beta_0 \lambda_{\mathrm{mfp}}
\equiv \lambda_{\mathrm{mfp}}/R_T$ was defined in the introduction.

\begin{figure}
\includegraphics[width = 0.45\textwidth]{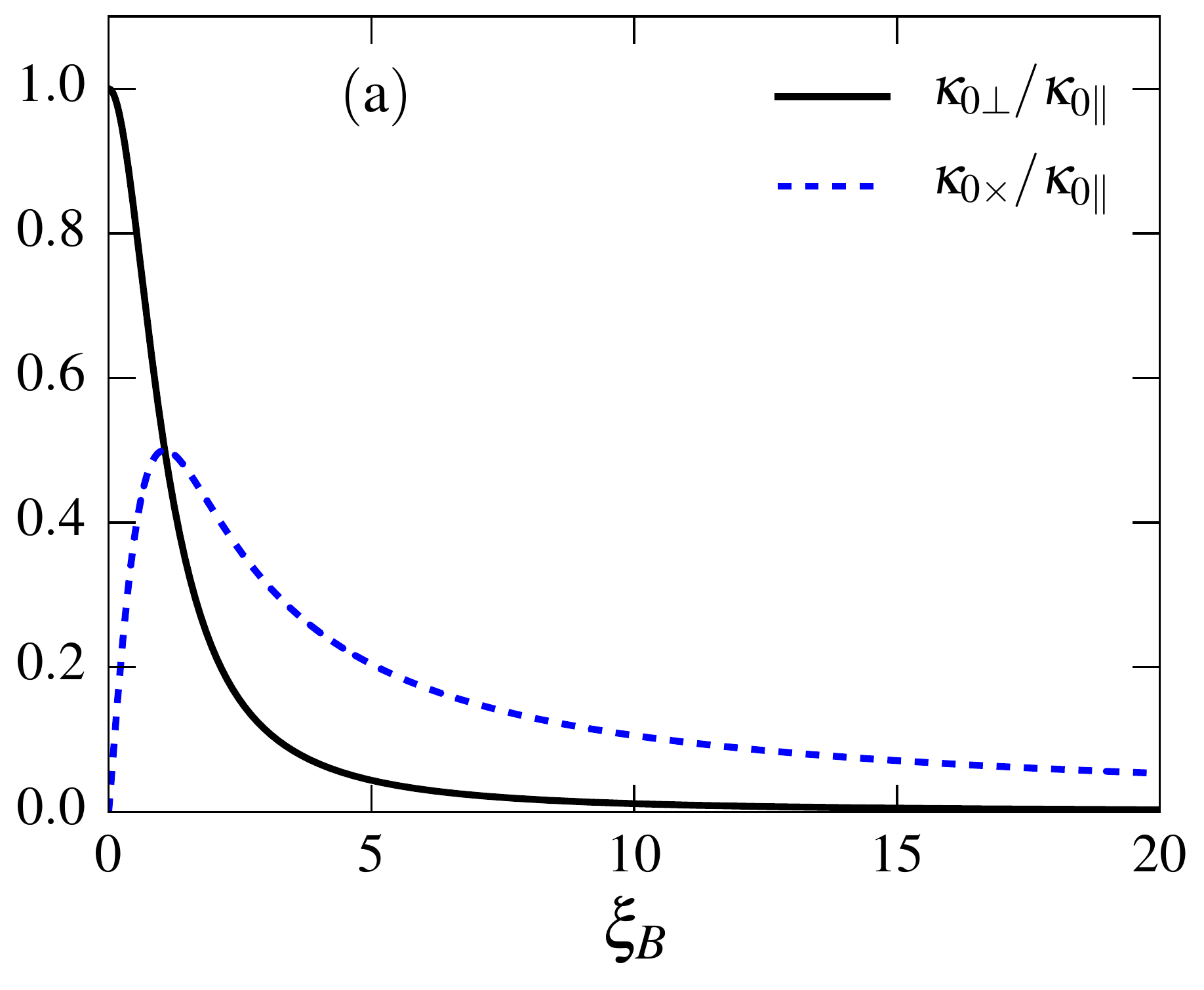}
\includegraphics[width = 0.45\textwidth]{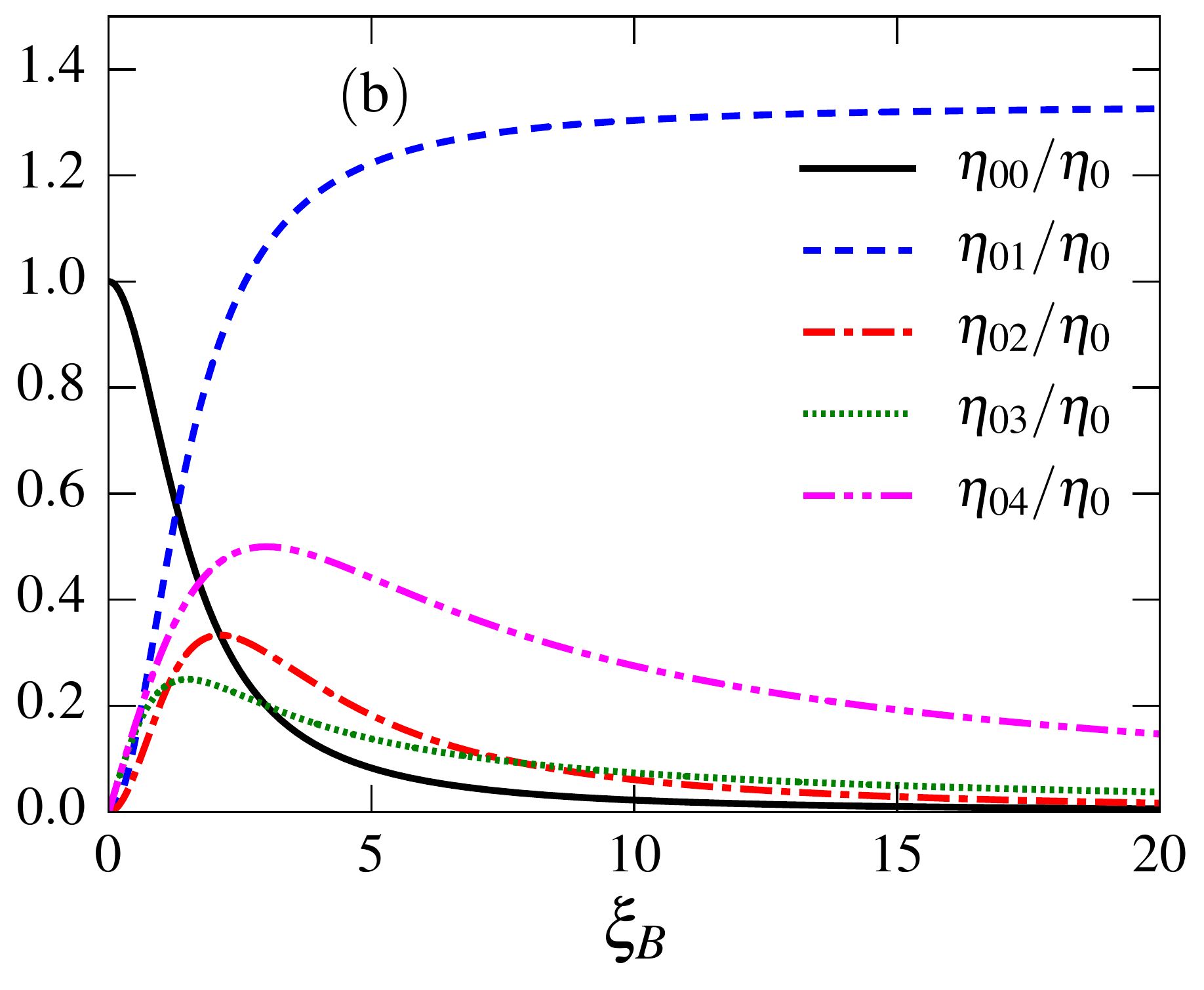}
 \caption{(Color online) The magnetic-field dependence of the diffusion coefficients (a) and the shear-viscosity 
 coefficients (b). }  \label{fig:coefficients}
\end{figure}

As expected, the longitudinal diffusion is solely given in terms of the mean free path, since the magnetic field 
does not affect the dynamics in the $b^{\mu}$ direction. On the other hand, there is an interplay between the mean 
free path and the thermal Larmor radius $R_T$ for the transverse diffusion, 
since the underlying particles not only collide but also undergo cyclotron motion.
The magnetic-field dependence of these coefficients is shown in Fig.~\ref{fig:coefficients} (a). 

Let us consider the limiting case where the mean free path is much larger than the thermal Larmor radius, i.e., 
$\xi_B \gg 1$. This can be achieved
either for fixed $B$ by decreasing the temperature or density, such that 
the mean free path increases, or by increasing
the magnetic field $B$, and thus decreasing the Larmor radius, for fixed
density, i.e., fixed mean free path. In this limit, 
\begin{equation}
\kappa _{0\parallel } = \frac{3 \lambda_{\mathrm{mfp}}n_{f0} }{16}\;, \;\;\;
\kappa _{0\perp } \simeq \frac{16}{75}\frac{\lambda_{\mathrm{mfp}}n_{f0} }{\xi_B^2}\;, \;\;\;
\kappa _{0\times } \simeq \frac{\lambda_{\mathrm{mfp}}n_{f0}}{5 \xi_B} \equiv \frac{n_{f0} R_T}{5}\; . \label{97}
\end{equation}
As expected, the Hall diffusion coefficient $\kappa _{0\times }$ assumes a value
which is independent of the mean free path. Note, however, that we obtain a non-zero value for this quantity.
The unique relationship between the diffusion coefficient and the electric conductivity (the Wiedemann-Franz law
mentioned above) then implies that also the Hall conductivity is non-zero. 
This result is different from the vanishing value quoted in
Eq.\ (8.198) of Ref.\ \cite{Cercignani_book}, valid for
a mixture of an ultrarelativistic electron gas and a non-relativistic ion gas.

Finally, inserting Eq.\ (\ref{rho_NS_mu_nu}) into Eq.\ (\ref{asy_rho_mu_nu}) leads to the following set of equations
for the shear-viscosity coefficients,
\begin{eqnarray}
\sum_{n=0}^{N_{2}}\left( \mathcal{A}_{rn}^{\left( \ell \right) }\eta
_{n0}+4\,\textswab{q}B\mathcal{F}_{1-r,n}^{\left( 2\right) }\eta _{n3}\right) 
&=&\alpha _{r}^{\left( 2\right) }\;, \\
\sum_{n=0}^{N_{2}}\left( \mathcal{A}_{rn}^{\left( 2\right) }\eta _{n3}
-\textswab{q}B \mathcal{F}_{1-r,n}^{\left( 2\right) }\eta _{n0}\right)  &=&0\;, \\
\sum_{n=0}^{N_{2}}\left( \mathcal{A}_{rn}^{\left( 2\right) }\eta _{n4}
-\mathcal{A}_{rn}^{\left( 2\right) }\eta _{n3}-\textswab{q}B\mathcal{F}_{1-r,n}^{\left( 2\right) }\eta _{n2}\right)  &=&0\;, \\
\sum_{n=0}^{N_{2}}\left( \mathcal{A}_{rn}^{\left( 2\right) }\eta _{n2}
+\textswab{q}B \mathcal{F}_{1-r,n}^{\left( 2\right) }\eta _{n4}
-4\,\textswab{q}B\mathcal{F}_{1-r,n}^{\left( 2\right) }\eta _{n3}\right)  &=&0\;, \\
\sum_{n=0}^{N_{2}}\left( 3\mathcal{A}_{rn}^{\left( 2\right) }\eta
_{n1}-16\,\textswab{q}B\mathcal{F}_{1-r,n}^{\left( 2\right) }\eta _{n3}\right)  &=&0\;.
\end{eqnarray}
In the 14-moment approximation ($N_2=0$) the above set of equations is solved by
\begin{eqnarray}
\eta _{00} &=&\eta_r \left[1+4\left( \textswab{q}B
\frac{\mathcal{F}_{1-r,0}^{\left(2\right) }}{\mathcal{A}_{r0}^{\left( 2\right) }}\right) ^{2}\right]^{-1}\;, \\
\eta _{01} &=&\frac{16}{3}\left( \textswab{q}B\frac{\mathcal{F}_{1-r,0}^{\left(2\right) }}{
\mathcal{A}_{r0}^{\left( 2\right) }}\right) ^{2}\eta _{00}\;, \\
\eta _{02} &=& 3\left( \textswab{q}B\frac{\mathcal{F}_{1-r,0}^{\left( 2\right)}}{\mathcal{A}_{r0}^{\left( 2\right) }}\right) ^{2}
\left[1+\left( \textswab{q}B\frac{\mathcal{F}_{1-r,0}^{\left( 2\right) }}{\mathcal{A}_{r0}^{\left( 2\right) }}\right) ^{2}\right]^{-1}
\eta _{00}\;, \\
\eta _{03} &=& \textswab{q}B\frac{\mathcal{F}_{1-r,0}^{\left( 2\right) }}{\mathcal{A}_{r0}^{\left( 2\right) }}\, \eta _{00}\; , \\
\eta _{04} &= & \eta_r \, \textswab{q}B\frac{\mathcal{F}_{1-r,0}^{\left(2\right) }}{\mathcal{A}_{r0}^{\left( 2\right) }}
 \left[ 1+\left(\textswab{q}B\frac{\mathcal{F}_{1-r,0}^{\left( 2\right) }}{\mathcal{A}_{r0}^{\left(2\right) }}\right) ^{2}\right]^{-1}
\;.
\end{eqnarray}
where $\eta_r = \alpha _{r}^{\left( 2\right) }/\mathcal{A}_{r0}^{\left(2\right) }$ corresponds 
to the usual shear-viscosity coefficient.
As expected, when $B\rightarrow 0$, only $\eta_{00}$ remains non-zero, such that
$\eta_0^{\mu\nu\alpha\beta} \sim \Delta^{\mu \nu \alpha \beta}$, as expected. Note that, for $B \neq 0$, the
``standard'' shear-viscosity coefficient $\eta_{00}$ is reduced as compared to its value for $B=0$. This
reduction of viscosity is similar to the mechanism suggested in Ref.\ \cite{Asakawa:2006tc}, giving rise to the 
so-called ``anomalous viscosity'', although that work considered gluon instead of electromagnetic fields.

In the limit of a massless Boltzmann gas and for a constant cross section, we obtain for $r=0$
the quantities $\alpha_0^{(2)} = 4P_0/5$, $\mathcal{F}_{10}^{\left( 2\right) }=\beta _{0}/5$, 
and $\mathcal{A}_{00}^{\left(2\right) }=3/(5\lambda _{\mathrm{mfp}})$. This yields
$\eta _{0}=4\lambda _{\mathrm{mfp}}P_0/3$  and
\begin{equation}
\eta _{00} = \frac{12 \lambda_{\mathrm{mfp}}P_0 }{9 + 4 \xi_B^2}\;,\;\;\;
\eta_{01} = \frac{64}{9} \, \frac{\xi_B^2 \lambda_{\mathrm{mfp}} P_0 }{
9 + 4 \xi_B^2}\;, \;\;\;
\eta_{02} =\frac{36\xi_B^2 \lambda_{\mathrm{mfp}} P_0 }{
[9 + 4 \xi_B^2][9 + \xi_B^2]}\;,\;\;\;
\eta _{03} = \frac{4 \xi_B \lambda_{\mathrm{mfp}} P_0 }{
9 + 4 \xi_B^2}\;,\;\;\;
\eta _{04} = \frac{4 \xi_B \lambda_{\mathrm{mfp}} P_0 }{
9 +  \xi_B^2}\;.
\end{equation}
The magnetic field dependence of these coefficients is shown in Fig.~\ref{fig:coefficients}(b). 
For a large ratio of mean free path to thermal Larmor radius, $\xi_B\gg 1$,
\begin{displaymath}
\eta _{00} = \frac{1}{3}\eta_{02} \simeq \frac{9}{4}\frac{\eta_0}{\xi_B^2}\;,\;\;\;
\eta_{01} \simeq \frac{4}{3} \eta_0\;, \;\;\;
\eta _{03} = \frac{1}{4}\eta_{04} \simeq \frac{\lambda_{\mathrm{mfp}}P_0 }{\xi_B} \equiv P_0 R_T\;.
\end{displaymath}
In this limit, the last two viscosities, $\eta_{03}$ and $\eta_{04}$, become independent of $\lambda_{\mathrm{mfp}}$. 
They appear purely due to the Lorentz force (and are thus named Hall viscosities). 
The relation $\eta _{03} = \eta_{04}/4$ holds also in the
non-relativistic case~\cite{Lifshitzbook}. We note that a similar study of the shear-viscosity coefficients in
the Navier-Stokes limit was recently performed in Ref.\ \cite{Mohanty:2018eja}, using the Boltzmann equation 
in the relaxation-time approximation.

Finally, we remark that the effect of a magnetic field on the shear viscosity of a strongly coupled $\mathcal{N}=4$ 
supersymmetric 
Yang-Mills plasma with a large number of colors was studied in Ref.\ \cite{Critelli:2014kra}. In this case, it was shown that 
the ratio between $\eta_{00}$ and the entropy density $s$ does not change with the magnetic field, 
$\eta_{00}/s = 1/(4\pi)$, while the ratio $(\eta_{00}+\eta_{02})/s$, considered in Ref.\ \cite{Critelli:2014kra}, was found to 
be suppressed in strong magnetic fields. This illustrates how the microscopic assumptions regarding the fluid, i.e., strong 
versus weak coupling, may alter its response to magnetic fields.

\subsection{Second-order magnetohydrodynamical equations of motion}
\label{sec:diss_eom_2ndorder}

We now derive the equations of motion for non-resistive, second-order dissipative magnetohydrodynamics. 
In this case, all terms in Eqs.\
(\ref{D_rho}) -- (\ref{D_rho_munu}) are kept, but the irreducible moments $\rho_r^{\mu_1 \cdots \mu_\ell}$ with
$r \neq 0$ are replaced using the 14-moment approximation ($N_0 = 2,\, N_1=1,\, N_2 =0$) using 
Eq.\ (\ref{useful}). With the definitions (\ref{kinetic:rho_0}) -- (\ref{kinetic:rho_0_munu}) we obtain 
\begin{align}
\rho _{r}& = \sum_{n=0,\neq 1,2}^{N_{0}}\rho _{n}\mathcal{F}_{-r,n}^{\left( 0\right) }
=-\frac{3}{m_{0}^{2}}\Pi 
\frac{J_{r0}D_{30}+J_{r+1,0}G_{23}+J_{r+2,0}D_{20}}{J_{20}D_{20}+J_{30}G_{12}+J_{40}D_{10}}\;,
\label{OMG_rho} \\
\rho _{r}^{\mu }& = \sum_{n=0,\neq 1}^{N_{1}}\rho _{n}^{\mu }\mathcal{F}_{-r,n}^{\left( 1\right) }
=V_f^{\mu } \frac{J_{r+2,1}J_{41}-J_{r+3,1}J_{31}}{D_{31}} \;,
\label{OMG_rho_mu} \\
\rho _{r}^{\mu \nu }& = \sum_{n=0}^{N_{2}}\rho _{n}^{\mu \nu }\mathcal{F}_{-r,n}^{\left( 2\right) }
=\pi ^{\mu \nu }\frac{J_{r+4,2}}{J_{42}}\;,
\label{OMG_rho_mu_nu}
\end{align}
while all higher-rank tensors ($\ell > 2$) are assumed to vanish. The above formulas also hold for negative values of $r$.

For $r= 0$ Eq.\ (\ref{D_rho}), together with Eqs.\ (\ref{OMG_rho}) -- (\ref{OMG_rho_mu_nu}), 
leads to an equation of motion for the bulk viscous pressure
\begin{equation}
\tau _{\Pi }\dot{\Pi}+\Pi =-\zeta \theta -\ell _{\Pi V}\, \nabla_{\mu}V_f^{\mu } -\tau _{\Pi V}\, V_f^{\mu }\dot{u}_{\mu }
-\delta _{\Pi \Pi }\, \Pi \theta  -\lambda _{\Pi V}\, V_f^{\mu }\nabla_{\mu }\alpha _{0} +\lambda _{\Pi \pi }\, \pi^{\mu \nu }
\sigma _{\mu \nu }\; .  \label{relax_bulk}
\end{equation}
Similarly, taking $r=0$ we obtain a relaxation equation for the particle diffusion current
from Eq.\ (\ref{D_rho_mu}) 
\begin{align}
\tau _{V}\dot{V}_f^{\left\langle \mu \right\rangle }+V_f^{\mu }& =\kappa
\nabla ^{\mu }\alpha_0 - \tau_V V_{f, \nu }\omega ^{\nu \mu }-\delta_{VV}\,V_f^{\mu }\theta -\ell _{V\Pi }\nabla ^{\mu }\Pi 
+\ell _{V\pi }\Delta^{\mu \nu }\nabla _{\lambda }\pi _{\nu }^{\lambda }+\tau _{V\Pi }\,\Pi \dot{u}^{\mu }
-\tau _{V\pi }\,\pi ^{\mu \nu }\dot{u}_{\nu }  \notag \\
& -\lambda _{VV}\,V_{f,\nu }\sigma ^{\mu \nu }+\lambda _{V\Pi }\,\Pi \nabla^{\mu }\alpha _{0}
-\lambda _{V\pi }\, \pi ^{\mu \nu }\nabla _{\nu }\alpha _{0} \notag \\
& -\delta _{VB}\, \textswab{q}Bb^{\mu \nu }V_{f,\nu }\;,  \label{relax_heat}
\end{align}
The relaxation equation of the shear-stress tensor follows from Eq.\ (\ref{D_rho_munu}) for $r=0$,
\begin{align}
\tau _{\pi }\dot{\pi}^{\left\langle \mu \nu \right\rangle }+\pi ^{\mu \nu }& =2\eta \sigma ^{\mu \nu }
+2\tau_{\pi} \pi _{\lambda }^{\left\langle \mu \right. }\omega^{\left. \nu \right\rangle \lambda }
-\delta _{\pi \pi }\,\pi ^{\mu \nu }\theta 
-\tau _{\pi \pi }\,\pi ^{\lambda \left\langle \mu \right. }\sigma _{\lambda}^{\left. \nu \right\rangle }
+\lambda _{\pi \Pi }\, \Pi \sigma ^{\mu \nu } \notag \\
& -\tau _{\pi V}\, V_f^{\left\langle \mu \right. }\dot{u}^{\left. \nu\right\rangle }
+\ell _{\pi V} \nabla ^{\left\langle \mu \right.}V_f^{\left. \nu \right\rangle }
+\lambda _{\pi V}\, V_f^{\left\langle \mu\right. }\nabla ^{\left. \nu \right\rangle }\alpha _{0}  \notag \\
& -\delta _{\pi B}\, \textswab{q} B b^{\alpha \beta }\Delta _{\alpha \kappa }^{\mu \nu}
g_{\lambda \beta }\pi ^{\kappa \lambda }\;.  \label{relax_shear}
\end{align}
The coefficients of the terms without explicit dependence on the magnetic field are
given in Appendix C of Ref.\ \cite{Denicol:2012cn} (note that
$n^{\mu }\leftrightarrow V_f^{\mu }$ and the index $n \leftrightarrow V$). 
In deriving these equations of motion only the linear contributions arising from the collision integrals were retained. 
We remark that, given our assumptions, 
the omitted nonlinear terms display no dependence on the magnetic field and were already calculated 
in Ref.\ \cite{Molnar:2013lta}.

To the best of our knowledge, Eqs.\ (\ref{relax_bulk}), (\ref{relax_heat}), and (\ref{relax_shear}) provide the first 
formulation of non-resistive, second-order dissipative magnetohydrodynamics that can be causal 
and linearly stable around 
equilibrium, in contrast to the Navier-Stokes approximation derived in Sec.\ \ref{sec:diss_eom_NS}. As such, this new 
system of equations is suitable to investigate the effects of magnetic fields on relativistic dissipative fluid dynamics, e.g.\
in heavy-ion collisions.

The coefficient of the term involving the magnetic field in Eq.\ (\ref{relax_heat}) is 
\begin{equation}
\delta _{VB}=\frac{ \mathcal{F}_{10}^{(1)}+ \alpha _{0}^{h}}{\mathcal{A}_{00}^{\left( 1\right) }}\; ,
\end{equation}
while the corresponding coefficient in Eq.\ (\ref{relax_shear}) is
\begin{equation}
\delta _{\pi B}=2\, \frac{\mathcal{F}_{10}^{(2)}}{\mathcal{A}_{00}^{\left( 2\right) }}\;.
\end{equation}

In the limit of a massless Boltzmann gas with constant cross section, 
$\alpha_0^h=- \beta_0/4$, $\mathcal{F}_{10}^{(1)}= 2\beta_0 /3$,
$\mathcal{F}_{10}^{(2)} = \beta_0 /5$,
$\mathcal{A}_{00}^{(1)} = 4/(9 \lambda_{\mathrm{mfp}})$, and
$\mathcal{A}_{00}^{(2)} = 3/(5 \lambda_{\mathrm{mfp}})$, such that
\begin{equation}
\delta _{VB}
=\frac{15}{16} \, \beta_0 \lambda_{\mathrm{mfp}}\; ,\;\;\;
\delta _{\pi  B}=\frac{2}{3}\, \beta_0 \lambda_{\mathrm{mfp}}\;.
\end{equation}

Let us finally comment on the first-order Navier-Stokes limit of the second-order equations 
(\ref{relax_bulk}) -- (\ref{relax_shear}).
Note that the first terms on the right-hand sides, proportional to the standard bulk- and shear-viscosity as well as 
particle-diffusion coefficients, are actually independent of the magnetic field. But these are not the only first-order
terms in these equations: without an assumption about the magnitude of the magnetic field,
also the last terms in Eqs.\ (\ref{relax_heat}), (\ref{relax_shear}) are formally of first order in a small quantity ($V_{f,\nu}$
or $\pi^{\kappa \lambda}$, respectively).
As demonstrated in the previous Sec.\ \ref{sec:diss_eom_NS}, these terms are to be combined with the
first-order terms on the left-hand side and, after inversion of the respective coefficient matrices, then lead to 
the various new anisotropic transport coefficients discussed above.

On the other hand, when solving the second-order equations (\ref{relax_heat}) and (\ref{relax_shear}), one
does not need to replace the standard viscosity and
particle-diffusion coefficients with the new anisotropic transport coefficients found in Sec.\ \ref{sec:diss_eom_NS},
because the effect of the magnetic field is already taken into account by the terms $\sim B$ in
these equations.

\section{Conclusions and outlook}
\label{conclusions}

We have derived, for the first time, the equations of motion for non-resistive, second-order 
dissipative magnetohydrodynamics from the Boltzmann equation. The
derivation is based on the moment expansion of the Boltzmann equation
coupled to a magnetic field for a single-component gas of 
particles without dipole moment or spin. The magnetohydrodynamical equations of
motion were obtained in the 14-moment approximation. This is essentially a generalization of
Israel-Stewart fluid dynamics to the case of a non-vanishing magnetic field.
Despite our simplifying assumptions, the results exhibit the basic
structure of second-order dissipative magnetohydrodynamics, in particular how
the magnetic field couples to the dynamical evolution of the
dissipative quantities. In particular, we note that within our
approximations the form of the equations remains close to that of
Israel-Stewart theory, with additional terms that couple the fluid to the magnetic
field. As such, the new set of second-order dissipative magnetohydrodynamical 
equations derived here allows one to investigate the effects 
of magnetic fields in relativistic dissipative fluids in a causal and linearly stable manner. Moreover, we have shown how 
the first-order transport coefficients split into several components,
recovering the results of Refs.~\cite{Huang:2011dc,Hernandez:2017mch}, with the 
notable difference that there is only one bulk-viscosity coefficient in our 
approximation. The reason for this is our assumption that the collision
integral is independent of the magnetic field. 

There are many possible directions for future work:
(i) The 14-moment approximation gives only an estimate for the values of the transport
coefficients. Improved values can be obtained by resumming higher orders in $N_\ell$ in the moment
expansion, as demonstrated in Ref.~\cite{Denicol:2012cn}. 
(ii) Resistive, second-order dissipative magnetohydrodynamics is obtained by keeping the electric field
$E^\mu$ in the equations of motion. 
(iii) An extension to spin degrees of freedom allows to include effects of polarization and magnetization
\cite{Israel:1978up}.
(iv) A relativistic treatment requires to take into account antiparticles with opposite electric charge.
These and further questions will be addressed in future work. 

\begin{acknowledgments}

The authors would like to thank T.\ Lappi for pointing out the similarity of the reduction of dissipative
transport coefficients in a magnetic field observed here to the mechanism suggested in Ref.\ \cite{Asakawa:2006tc}.
G.S.D.\ greatly acknowledges the warm hospitality of the Department of Physics of Goethe University,
where part of this work was done.
E.M.\ and D.H.R.\ greatly acknowledge the warm hospitality of the Department of
Physics of the University of Jyv\"askyl\"a, where part of this work was done.
This work was supported by the Collaborative Research Center CRC-TR 211 ``Strong-interaction matter
under extreme conditions'' funded by DFG.
G.S.D.\ and J.N.\ thank Conselho Nacional de Desenvolvimento Cient\'ifico e Tecnol\'ogico (CNPq) for financial support. 
X.G.H.\ is supported by the Young 1000 Talents Program of China, NSFC with Grant No.~11535012 and No.~11675041.
E.M.\ is supported by the Bundesministerium f\"ur Bildung und Forschung (BMBF) and by the Research 
Council of Norway, (NFR) Project No.~255253/F50.
H.N.\ is supported by the European Union's Horizon 2020 research and innovation
programme under the Marie Sklodowska-Curie grant agreement no.\ 655285 and by the
Academy of Finland, project 297058.
J.N.\ and G.M.M.\ thank Funda\c c\~ao de Amparo \`a Pesquisa do Estado de S\~ao Paulo (FAPESP) under grants  
2015/50266-2 (2017/05685-2) and 2016/13517-0, respectively, for financial support.
D.H.R.\ is partially supported by the High-end Foreign Experts project GDW20167100136 of the State
Administration of Foreign Experts Affairs of China. 
\end{acknowledgments}

\appendix
\section*{Appendix}

Our conventions for the rank-four Levi-Civit\`a tensor $\epsilon^{\mu \nu \alpha \beta }$ are as follows.
We take $\epsilon^{0123}=+1$, which implies $\epsilon ^{\mu \nu \alpha \beta }=-\epsilon _{\mu\nu \alpha \beta }$.
We also have the relations
\begin{equation} \label{aux1}
\epsilon^{\mu \alpha \beta \gamma} \epsilon_{\nu \alpha \rho \sigma} =
\delta^\mu_\nu \left( \delta^\beta_\sigma \delta^\gamma_\rho - \delta^\beta_\rho \delta^\gamma_\sigma \right)
+ \delta^\mu_\rho \left( \delta^\beta_\nu \delta^\gamma_\sigma - \delta^\beta_\sigma \delta^\gamma_\nu \right)
+ \delta^\mu_\sigma \left( \delta^\beta_\rho \delta^\gamma_\nu - \delta^\beta_\nu \delta^\gamma_\rho \right)\;,
\end{equation}
and 
\begin{equation} \label{aux2}
\epsilon ^{\mu \nu \alpha \beta }\epsilon_{\kappa \lambda \alpha \beta }
=2\left( \delta_{\lambda }^{\mu }\delta_{\kappa}^{\nu } -\delta_{\kappa}^{\mu }\delta_{\lambda}^{\nu }\right) \;.
\end{equation}
In flat Minkowski space, all Kronecker deltas can be replaced by the mixed contra- and covariant metric tensor, e.g.\
$\delta^\mu_\nu \equiv g^\mu_\nu$.

\end{document}